\documentclass[useAMS,usegraphicx,usenatbib]{mn2e}

\usepackage{enumerate}

\title[Globular Clusters in M104]{Diamonds on the Hat:  Globular Clusters in 
The Sombrero Galaxy (M104)$^1$}
\author[Harris et al.]{Wlliam~E.~Harris$^1$, Lee~R.~Spitler$^2$, 
Duncan~A.~Forbes$^2$, and Jeremy~Bailin$^1$ \\
$^{1}$Department of Physics \& Astronomy, McMaster University,
      Hamilton ON L8S 4M1, Canada \\
$^{2}$Centre for Astrophysics and Supercomputing, Swinburne University
      of Technology, Hawthorn, Victoria 3122, Australia }

\begin{document}
\maketitle
\label{firstpage}
\begin{abstract}
Images from the Hubble Space Telescope Advanced Camera for Surveys 
are used to carry out a new photometric study of the globular clusters
(GCs) in M104, the Sombrero galaxy.
The primary focus of our study is the characteristic 
distribution function of linear sizes (SDF) of the GCs. We 
measure the effective radii for 652 clusters with
PSF-convolved King and Wilson dynamical model fits.
The SDF is remarkably similar to those measured
for other large galaxies of all types, adding strong support to the
view that it is a ``universal'' feature of globular cluster systems.

We use the Sombrero and Milky Way data and the formation models
of Baumgardt \& Kroupa (2007) to develop a more general interpretation
of the size distribution function for globular clusters.
We propose that the shape of the SDF that we see today for GCs 
is strongly influenced by the early rapid mass loss
during their star forming stage, coupled with stochastic differences
from cluster to cluster in the star formation efficiency (SFE)
and their initial sizes.
We find that the observed SDF shape can be accurately predicted by
a simple model in which the protocluster clouds had
characteristic sizes of $0.9 \pm 0.1$ pc and SFEs of $0.3 \pm 0.07$.

The colors and luminosities of the M104 clusters show the clearly
defined classic bimodal form.  The blue sequence exhibits a
mass/metallicity relation (MMR), following a scaling of
heavy-element abundance with luminosity of $Z \sim L^{0.3}$
very similar to what has been found in most giant elliptical
galaxies.  A quantitative self-enrichment model provides
a good first-order match to the data for the same initial
SFE and protocluster size that were required to explain the SDF.

We also discuss various forms of the globular cluster Fundamental
Plane (FP) of structural parameters, and show that useful tests
of it can be extended to galaxies beyond the Local Group.  The M104
clusters strongly resemble those of the Milky Way and other 
nearby systems in terms of such test quantities as integrated surface density
and binding energy.
\end{abstract}


\section{Introduction}
\footnotetext[1]{This work was based on observations with the 
NASA/ESA Hubble Space Telescope, obtained at the Space Telescope 
Science Institute, which is operated by the Association of 
Universities for Research in Astronomy, Inc.,under NASA contract 
NAS 5-26555. }

Our knowledge of globular cluster (GC) systems in spiral
and disk galaxies is far more limited than those
in elliptical galaxies, and consists primarily of the samples 
in the Milky Way and in M31, along with 
a handful of more distant disk galaxies 
\citep{kisslerpatig1999,goudfrooij2003, chandar2004, rhode2007, 
spitler2006,mora2007,can07,deg07}.
These should be compared with the
many extensive studies of the GC systems in elliptical
galaxies, particularly the giant E's in which GCs can
be found by the thousands 
\citep[e.g.][to name only a few]{larsen01,brodie06,peng06,harris09}.   
But if we are to probe the systematic properties of the GCs themselves and
to understand further the common nature of these classically
ancient objects, we need to investigate them closely across
all types of host galaxies.

The Sombrero galaxy (M104 = NGC 4594), a giant Sa system, is the disk galaxy with the
largest known population of globular clusters (GCs) and for
that reason alone is 
a particularly attractive system.  The fact that it is
distinctly closer ($d=9$ Mpc) than the
Virgo cluster, and also nearly edge-on to our line of sight so that
our view of its GC population is minimally affected by issues of
reddening and contamination from the disk itself, makes M104
an almost unique target. \citet{rhode2004}, from wide-field
ground-based imaging, derive a total GC population of 1900 clusters.
They find that the spatial extent of the system is at least as large as 40 kpc
in projected galactocentric distance.

In an earlier study \citep[][which we refer to as Paper I]{spitler2006},
a first round of HST-based photometry of M104 GCs and measurements of their scale
sizes was presented.  \citet{chandar2007} used the same material to evaluate
the trend of mean GC density with galactocentric distance.
The raw data consisted of
a special imaging dataset of six fields taken with the ACS Wide Field Channel
as part of a Hubble Heritage Project on M104 (PI: K.~Noll,
PID 9714). This mosaic was observed in the
standard $BVR$ filters (F435W, F555W, F625W) and covers a total field
size of about $600'' \times 400''$ centered on M104.  In Paper I, photometry 
from the individual images was used to
discuss the distribution of the GCs in luminosity and color, their spatial
distributions around the galaxy, the differences between the red (metal-rich)
and blue (metal-poor) GC subsystems, and the correlations of 
cluster sizes with GC luminosity and galactocentric distance.  In addition,
Paper I also provided the first indication that the metal-poor GCs in disk galaxies showed
a correlation between mass and metallicity that had already been uncovered
for elliptical galaxies.

In the present study, we use the reprocessed mosaic of ACS images
to measure the GCs in M104.  A particular emphasis of this study is
the use of newer algorithms to take a deeper look into the \emph{intrinsic
scale sizes} of the GCs, and carry out a
more extensive set of comparisons with other galaxies.
This work leads us into a physical model for the origin of the 
GC size distribution that we observe today.

In Section 2 we describe in detail the photometry and size measurements.
In Section 3, we discuss the correlations of GC scale size with luminosity,
metallicity, and galactocentric distance, and show that versions of the
``Fundamental Plane'' of structural parameters can be accurately constructed
for GCs in galaxies this distant.  In Section 4, we introduce a quantitative model for
the physical origin of the GC size distribution. The
present-day GC size must be the result, not only of the slow
dynamical evolution of the cluster in the tidal field of the
galaxy, but also
of the expansion of the protoclusters during 
their early phase of star formation and rapid gas expulsion.  The key parameters
in this model are the original protocluster size, the star formation efficiency,
and (just as importantly) the dispersions in these two quantities.
Finally, in Section 5 we redetermine the correlations of GC color (metallicity)
with luminosity for both the bimodal sequences with our new photometry.
We conclude more strongly than before that this galaxy has a mass/metallicity
relation quite similar to those in the giant ellipticals.  
Section 6 gives a brief summary of our findings.
(Readers interested primarily in the results may skip directly to Section 3
without loss of continuity.)

Throughout this paper, we adopt the distance modulus $(m-M)_0 = 29.77$ ($d=9.0$ Mpc;
see Table 4 in Paper I) along with a foreground reddening $E_{B-V} = 0.05$
from the NASA/IPAC Extragalactic Database (NED).
At the adopted distance, the image scale is 1 pixel $= 0.05'' = 2.2$ parsecs.

\section{Goals of the Study}

The most prominent focus of the present discussion
is the characteristic
scale size of the GCs (the effective or half-light radius, denoted $r_h$).
We refer to the \emph{distribution function} of their sizes (effective radii) as the SDF.
At the 9-Mpc distance of M104, a GC with a typical size
$r_h \simeq 3$ pc subtends an angular diameter of about $0.14''$,
which is similar to the ACS/WFC point-spread function width
of FWHM $0.1''$.  In the sense defined by \citet{harris09}, the clusters
are therefore in the \emph{partially resolved} regime: their individual
profiles are easily distinguished from stars, and their
effective radii are clearly measurable.  Extensive tests of GC profile-fitting
for galaxies ranging out to distances of $\sim 50$ Mpc show that GC    
effective radii can be accurately measured as long as the stellar PSF
is precisely known and the cluster size $r_h$ is no smaller than about
10\% of the PSF fwhm \citep{kun98,lar99,jor05,georgiev08,harris09}.
As will be seen below, the clusters in M104 are far above this resolution limit.

We use our new measurements to establish
the SDF more accurately, and to define the ``Fundamental Plane'' of GC
structural parameters in comparison with other galaxies.
The half-\emph{mass} radius $r_{h,m}$ is an interesting and
important quantity for GC dynamics because it remains relatively constant
over many internal relaxation times \citep[e.g.][]{spi72,aar98,baum02,tre07}
and thus represents an intrinsic scale size that was established closer to its
formation time.  As we will discuss in Section 4 below, the physical origin
of the SDF is likely to be sensitively dependent on a combination of the star formation efficiency
(SFE) and the original scale size ($r(init)$) of the protoclusters.  
The fact that the SDF has
a near-universal form among globular cluster systems can then
be turned around to place surprisingly close limits on the SFE and $r(init)$.

The HST cameras, beginning with WFPC2, have provided the key instrumental
tools that have allowed us to resolve GCs in galaxies well beyond
the Local Group and to measure their scale sizes accurately.
Pioneering work of this type began a decade ago \citep{kun98,kun01,kun99,lar99,larsen01}
and the changeover from WFPC2 to the ACS camera soon led to faster gains and large
databases of cluster sizes \citep[e.g.][]{jor05,jor09,harris09}.
These surveys of resolved GC populations  
have started to reveal that $r_h$ also depends weakly
on three other cluster properties.
First of these is a gradual increase in mean $r_h$  with \emph{galactocentric distance}
\citep{hodge62,mateo87,vdb91,jor05,spitler2006,can07,gom07,harris09,madrid09}.
This trend is generally interpreted as the result of the weaker external
tidal field of the host galaxy with increasing distance 
\citep[see][]{vonh57,king62,vdb91,mur92,hp94,baum03},
raising the intriguing possibility that the GC sizes can provide a tracer
of the galaxy potential well independently of other approaches.

Secondly, in most galaxies a detectable difference in mean size with
cluster \emph{metallicity} has been found, with the red GCs being typically
15-20\% smaller than the blue ones 
\citep[][and Paper I]{kun98,kun99,larsen01,jor05,gom07,harris09}.
This second trend has been variously interpreted as (a) a geometric-projection
byproduct of the fact that the metal-richer clusters usually lie in
a more centrally concentrated spatial distribution than the metal-poor ones,
coupled with the growth in mean $r_h$ with $R_{gc}$  \citep{lar03};
(b) a result of metallicity-dependent stellar evolution times and
dynamical mass segregation within the clusters \citep{jor04}; or
(c) a residual of the cluster formation epoch, in which the gas within
a metal-richer protocluster could cool and collapse further 
than in a metal-poor one before forming stars \citep{harris09}.
Combinations of these effects are of course not ruled out.  

A third empirically established trend is the correlation of GC scale
size with cluster \emph{luminosity} or mass.  Even though there 
is obvious cluster-to-cluster scatter in intrinsic sizes,
their median scale size differs
little with absolute magnitude from a global value of $2.5- 3$ pc
\citep[e.g.][]{jor05,bar07}.  
It is only at the high-luminosity
end ($\sim 5 \times 10^5$ Solar masses and above) that a trend towards somewhat
larger $r_h$ seems to emerge.  This \emph{mass/radius relation} may,
possibly, bridge the sequence of normal GCs to other kinds of compact
stellar systems such as UCDs (Ultra-Compact Dwarfs) that have scale sizes of typically
$\sim 10 - 30$ pc \citep{has05,kis06,bar07,evs08,harris09}.  These somewhat
more massive compact systems \citep[variously called UCDs, Dwarf-Globular Transition
Objects (DGTOs), or Intermediate-Mass Objects (IMOs); see][]{for08}
obey a different scaling rule $r_h \sim M^{0.8}$ or higher 
\citep[see][]{for08,evs08}.
Globular clusters luminous enough to lie
in this intriguing transition region of $10^6 - 10^7 M_{\odot}$ 
are so rare that it is not yet clear whether there is
a ``universal'' mass/radius relation for them, or what its detailed form is.

For all the reasons outlined above, high-quality measurements of GC
scale sizes and structural parameters are of great value.  Because of
its proximity and large GC population, M104 provides one
of the few best opportunities for this kind of measurement outside
the Local Group.

\subsection{The Data and the Profile Models}

Preprocessing of the raw image data is
described in Paper I.  The mosaics of six pointings in each filter were constructed
at STScI and are publicly available.  Paper I also lays out in detail the selection
of the many hundreds of individual GCs in M104 field and rejection
of contaminating non-GCs scattered across the field.  The contaminants are
mostly small, faint background galaxies, but also include any starlike objects
($r_h =0$) according to the profile fits.
Selection also included the use of color, magnitude, and the distribution
of objects in a similar control field.  There are some 658 identified high-probability GCs
brighter than $R=24 (M_R \simeq -5.8)$ plus one UCD
in the final selected list, and we start the present paper with that list in hand.

In Paper I, the ISHAPE code of \citet{lar99} was used to determine
the GC effective radii through convolution of \citet{king62} model profiles
with the stellar PSF.  In each of the six original ACS/WFC fields
a single mean PSF was defined, and a profile with a central
concentration parameter $c = r_t/r_c = 30$ was adopted.
For each individual cluster the intrinsic profile width
$r_h$ (in practice, the fwhm) is varied until a best fit between the convolved
profile and the real GC is obtained.  

\begin{figure}
\includegraphics[angle=0,width=0.5\textwidth]{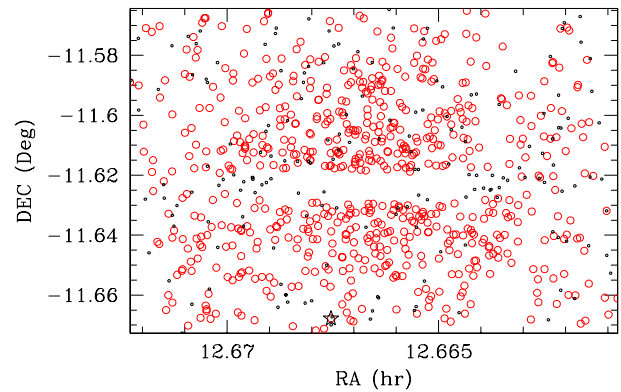}
\caption{Location of the target objects within M104 mosaic
field, which has dimensions $600'' \times 400''$.  The 659 GC candidates are
shown as open circles, while the 179 stars used for the PSFs are marked
as small dots.  The large star at bottom shows the location of the UCD
\citep{hau09}.
}
\label{xyplot}
\end{figure}

The ISHAPE code is a well tested
and fast method for obtaining useful $r_h$ distributions
for large samples of partially or marginally resolved objects 
\citep[see, e.g.][]{lar99,larsen01,georgiev08,harris09,madrid09}.
However, it does not easily handle situations where the PSF may
depend on position on the image, and  at a more fundamental level
it uses only analytic fitting functions to describe the GC profiles.
Although the \citet{king62} function was constructed to match real
GCs in the first place, it does not capture the full range of actual
profile shapes for GCs, UCDs, and related objects.

For the present work, we use the profile-fitting code of \citet{mcl08}.
This code has been applied previously to GCs in the Milky Way, the
Magellanic Clouds, M31, and NGC 5128 \citep{mcl05,bar07,mcl08}.
For the intrinsic GC profile, this algorithm can employ either the
\citet{king66} or \citet{wilson75} profiles, both of which are true
dynamical models constructed from simple but realistic assumptions about the energy
distribution function of the stars within the cluster.  In the discussion
below, we refer to these models as ``K66'' and ``W75''.\footnote{The
code can also use any of three analytic fitting functions including
the Sersic profile, a power-law profile, and the equally familiar
\citet{king62} analytic function, which we refer to as ``K62''.}
Although these
non-analytic models require a much larger investment in computing time,
their best-fit solutions 
lead immediately to physically relevant quantities such
as relaxation times, mean densities, binding energies, and fundamental-plane
parameters \citep{mcl05,mcl08}.  

A roughly similar code (KINGPHOT) has been used for
GCs in the Virgo galaxies 
\citep{jor05,jor09,peng09}.  Although the King model 
is quite familiar in the literature and describes most real GCs satisfactorily, 
the Wilson model has some distinct advantages
in its ability to handle real clusters
that have symmetric but \emph{radially extended} light profiles which
(within the confines of the King model) would nominally be thought to
have ``extra-tidal light''.  
As is extensively discussed by \citet{mcl05} and \citet{mcl08}, 
the key difference in the Wilson model is its smoother ramp-down 
of the stellar energy distribution
function towards higher energies, which can handle relatively larger numbers of stars
near the escape energy and a more extended envelope with
a nominally larger tidal radius.  

It is important to note that for small to moderate radii within
the cluster (that is, out to a few $r_h$), both the King and Wilson models
give very similar profile shapes and do equally well at fitting most real clusters.
These inner regions are populated by the low-energy stars far inside the 
tidal (escape) radius where the difference in the energy distribution
function between the two models is small.  For the clusters in the
nearby Local Group galaxies where the profiles can be thoroughly
sampled over large radii, the Wilson model often performs better, but 
\citep[as is shown in][]{mcl05,mcl08} there is no blanket ruling;
decisions are made on a cluster-by-cluster basis.

The Sombrero ACS mosaic field does, however, offer a particularly vivid case
where there is a strong difference between these two models.  
There is a single clearly identifiable luminous UCD belonging to the galaxy,
whose profile has an extended outer envelope which is accurately fit
by the Wilson model but where the King model completely fails.
We have discussed this special object in detail in \citet{hau09}.
In the present study, we apply both of these two dynamical model
profiles to the complete sample of GCs in the galaxy and explicitly
compare their quality of fit.

\subsection{Measurement of Cluster Profiles}

The profile-fitting code is the same one described fully in
\citet{mcl08}.  For each object, a stellar PSF (either an analytic function or
an empirical numeric profile) is convolved with the K66 or W75 model,
and the model parameters are adjusted until the convolution fits the observed
surface brightness profile of the cluster.  We carried out all the
measurements on the complete
wide-field mosaic for the entire field, rather than the six individual
fields as in Paper I.  The measurements were done from
smoothed (median-filtered) images 
in each filter with the large-scale light gradients removed.   
A $31 \times 31$ px box filter was used for the median filtering.
We then used \emph{stsdas/ellipse} to measure the radial profile
for each GC in our list, out to a radius determined by the point past
which the GC surface brightness fell well below the sky noise;
this limit ranged from $\ga 0.5''$ for the most luminous objects
to $0.1''$ for the very faintest ones.

To define the PSF profile that needs to be paired up with each GC, we chose here to
select the \emph{nearest star} to each individual GC, as we did in
\citet{hau09}.  In this way we circumvent trying to model the
small but potentially complex variation in PSF properties across
the complete mosaic of six ACS fields.  The stars are, in turn, empirically 
defined as those objects in our Paper I study that turned out to
have negligible intrinsic widths ($r_h = 0$).  We drew the PSF stars from
a candidate list of 179 stars with sufficiently high S/N, 
so the average projected separation between
any one GC and its accompanying PSF star is $\simeq 20''$.
The profiles of the stars were also measured through \emph{stdas/ellipse}
with the same parameters, and each GC/PSF pair was then run through
the fitting code to find the best-fit solutions.
The locations of the 658 target GCs and one UCD measured
previously, along with the 179 candidate PSF stars, are shown in
Figure \ref{xyplot}.  As is evident from the figure, we deliberately
avoid the parts of the galaxy projected on or near the plane of the disk, where
the profile fits would be badly compromised by complex background
light gradients and crowding.

\begin{figure}
\includegraphics[angle=0,width=0.5\textwidth]{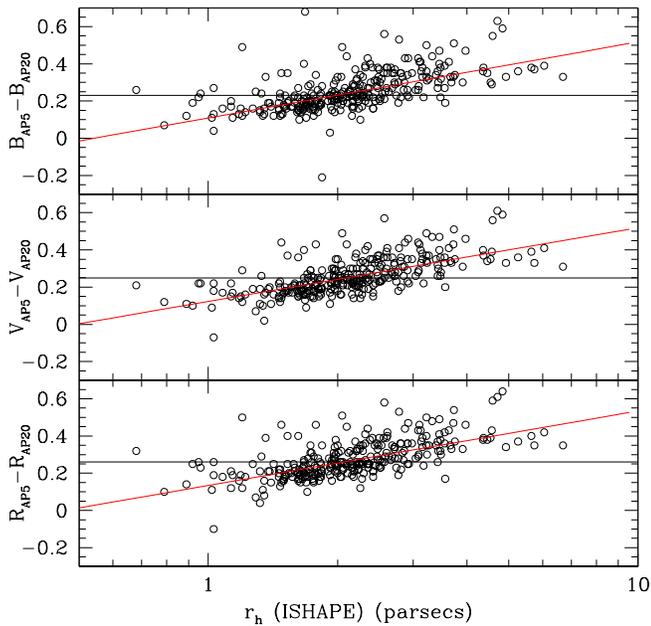}
\caption{Empirical aperture corrections to the cluster magnitudes.
The mean line through each set of points is given by the equation
in the text.
}
\label{apcorr}
\end{figure}

Lastly, we re-derive integrated magnitudes $B,V,R$ for all the GCs
with aperture photometry individually adjusted for cluster profile
width.  The approach we use is similar in principle to \citet{harris09}, where in
this case we start with a fixed-aperture magnitude measured through
5 px radius ($0.05''$) and then apply a magnitude correction 
to ``large'' radius (20 px) $\Delta m = m_5 - m_{20}$
that depends in turn on the cluster's scale size $r_h$.  
By direct comparison of the aperture magnitudes from ISHAPE through these two
apertures (Figure \ref{apcorr}), we find that $\Delta m$ increases
nearly linearly with log ($r_h$),
$$ \Delta m \, = \, 0.12 \, + \, 0.41 {\rm log} r_h $$
and does not differ significantly with bandpass.  Thus the aperture
corrections have no significant influence on the cluster colors
\citep[see also][]{harris09,jor09,peng09}.

\begin{table*}
\caption{Photometry and Structural Parameters for Globular Clusters in M104}
\label{datatab}
\begin{tabular}{lccccccccrr}
\hline
  ID  &   RA   &     DEC  & $R^{'}_{gc}$  &   $V$ & $B-V$ & $B-R$ &  $r_h$(pc) & $\pm$(pc) & $c$ & $\pm$ \\
\hline
   1 &  12.669456 &-11.639086 & 2.777 & 18.74 & 0.77 & 1.26 & 3.62 & 0.18 & 1.73 & 0.09 \\
   2 &  12.669275 &-11.642228 & 2.703 & 18.87 & 0.83 & 1.34 & 5.05 & 0.06 & 1.90 & 0.01 \\
   3 &  12.666692 &-11.602536 & 1.243 & 18.83 & 0.76 & 1.27 & 3.72 & 0.46 & 2.37 & 0.71 \\
   4 &  12.670719 &-11.641109 & 3.874 & 18.83 & 0.84 & 1.36 & 4.70 & 0.36 & 1.94 & 0.22 \\
   5 &  12.669567 &-11.610695 & 2.804 & 18.99 & 0.80 & 1.31 & 3.57 & 0.10 & 1.98 & 0.06 \\

\hline
\end{tabular}
\end{table*}

The output quantities from the profile fitting code that are of primary
interest here are the effective radius $r_h$ and its uncertainty
$\sigma(r_h)$; the central potential
parameter $W_0$ or equivalently the concentration $c$ and
its uncertainty ($c \equiv {\rm log} (r_t/r_c)$
where $r_t, r_c$ are the tidal and core radii); and a goodness
of fit $\chi^2$.  
As all other authors have found who have worked on GC size measurements
for distant galaxies \citep[e.g.][]{kun98,kun01,lar99,jor05,georgiev08,harris09},
we find that the central-concentration parameter $W_0$ or $c$ is
not very precisely determined from the best-fit solutions.  The reason
for this is simply that relatively small numbers of pixels are available for
the code to use for the profile fit relative to the PSF size.
In turn, some of the other structural
quantities calculated from $c$ and $r_h$ 
(such as the cluster core radius $r_c$, which is 
far smaller than the resolution limit imposed by the PSF) will not
be reliable.  Empirically, we can gauge how reliable our $c-$values are
by comparing the output values from the three different filters for the same object.  
These comparisons show that the internal consistency differs widely from
one cluster to another, but has a median uncertainty near $\sigma(c) = \pm 0.4$.

Fortunately, by far the most robust quantity in the solutions is the
effective radius itself.
As is discussed further elsewhere (see the references cited above), $r_h$ is
relatively insensitive to changes in $c$ in either direction because, for
GCs at distances in the $\sim 5-50$ Mpc range, it 
sits near what is essentially a ``best'' point in the profile:  it is
neither buried in the unresolved core of the cluster, nor lost in the noise 
blanketing its faint outermost envelope.  As long as the PSF profile
is accurately known, any
error in $c$ is compensated by adjustments in the total profile
shape to give the well determined radius enclosing half the light.
Fortunately, the M104 clusters
sit well above the empirical resolution limit of $r_h \sim 0.1 fwhm(PSF)$
below which any leverage on the structural parameters is lost.

\begin{figure}
\includegraphics[angle=0,width=0.5\textwidth]{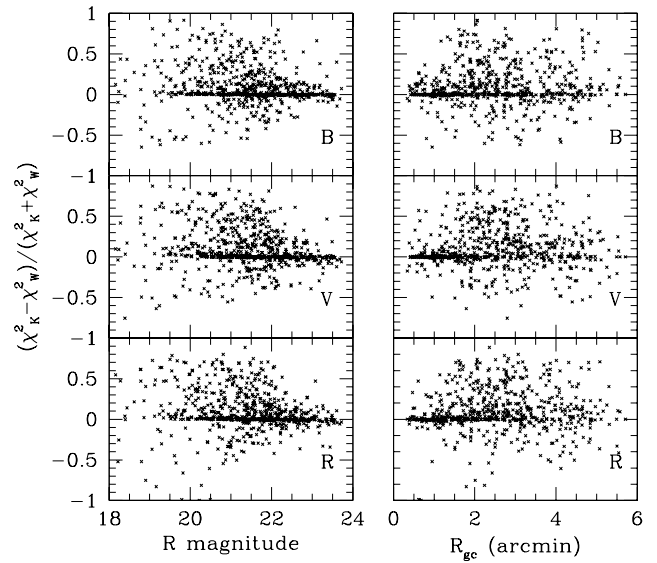}
\caption{\emph{Left panel:} Normalized $\chi^2$ ratio plotted versus
$R$ magnitude, for the three bandpasses $BVR$.  In this plot,
postive values indicate that the Wilson (1975) model is preferred to
the King (1966) model, while values near zero indicate that both
are equally good.  \emph{Right panel:}  Normalized $\chi^2$ ratio 
plotted against projected galactocentric distance.
}
\label{chifit}
\end{figure}

A direct comparison of the K66 and W75 models is shown
in Figure \ref{chifit}.  Here, as in \citet{mcl08} we use a normalized
$\chi^2$ ratio defined as $(\chi^2_K - \chi^2_W)/(\chi^2_K + \chi^2_W)$
where $\chi_K, \chi_W$ are the values from the King and Wilson fits
respectively.  For most of the clusters, the values of this ratio
scatter near zero, indicating that both models do about equally well
at matching the real profiles.  Overall, there is a slight
preponderance for the W75 models to match better (positive values
in the graphs) but no major differences or obvious trends with magnitude are evident.
By plotting the same goodness-of-fit ratio against projected galactocentric distance $R_{gc}$,
we test for any trends versus background sky noise (the large bulge
of M104 becomes considerably brighter at small $R_{gc}$), but
we find no systematic trends there either.

\begin{figure}
\includegraphics[angle=0,width=0.5\textwidth]{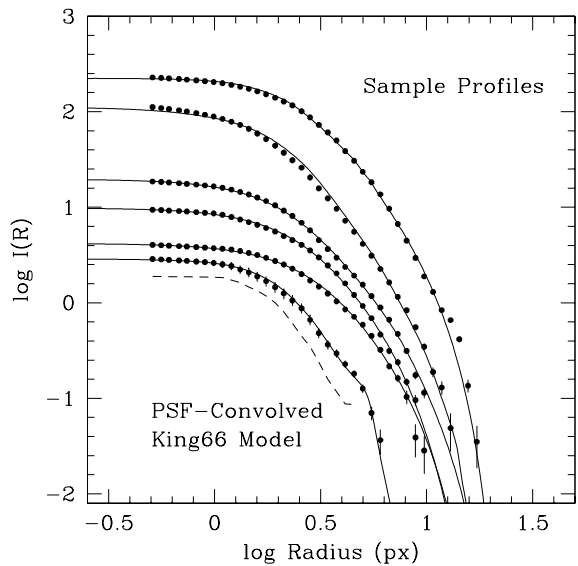}
\caption{Some sample profile fits with the King (1966) model
for six of the clusters, ranging from the brightest in the list
to the faintest (top to bottom).  Dots are the measured surface
brightnesses from \emph{stsdas/ellipse}, while solid lines
are the PSF-convolved K66 model solutions.  The dashed line
at bottom shows one of the PSF profiles.
}
\label{samplefits}
\end{figure}

Final successful profile fits and radius measurements for 652 clusters
were obtained.  Table \ref{datatab} lists the results giving in successive
columns a running ID number; right ascension and declination;
projected galactocentric distance in arcminutes; $V, B-V$, and $B-R$;
the mean $r_h$ over all three filters and its uncertainty;
and the mean King central concentration parameter $c$ and its internal
uncertainty.
The complete version of the table is available in the electronic edition.
In Figure \ref{samplefits}, we show several solutions to individual
clusters with the K66 model and drawn from the $R-$band filter.

\subsection{Measurement Uncertainties and Error Budget}

\begin{figure}
\includegraphics[angle=0,width=0.5\textwidth]{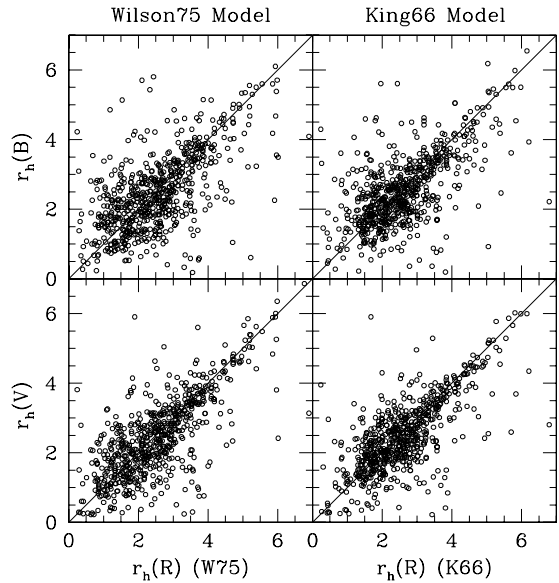}
\caption{\emph{Left panels:} Comparison between the cluster sizes
measured in the three different filters, as measured from the
Wilson 1975 model (upper panel is $B$ vs. $R$, lower panel is
$V$ vs. $R$; all scales are in parsec units).
\emph{Right panels:}  Same internal comparison for the measurements
with the King (1966) model.
}
\label{bvr}
\end{figure}

The measurement uncertainties on the $r_h$ values were evaluated
with a series of tests. First of these was the internal
consistency in the size measurement among the three filters.  
This comparison is shown in Figure \ref{bvr}, 
for both the K66 and W75 models.
In principle, the three filters give three independent 
measurements of the same quantity $r_h$ for the same cluster (and
for exactly the same PSF star),
with random scatter due only to the internal uncertainty of the fit.
We find good systematic agreement amongst the
filters; all three correlations scatter closely along the 1:1 lines,
so we make no systematic corrections between
filters.  However, the K66 model fits (right panels in the figure) show slightly
smaller scatter, fewer outliers, and thus higher internal consistency than the
W75 model.

In the end, we find little to choose between these two competing
models for most individual clusters.  Because of its 
slightly superior internal consistency
(Figure \ref{bvr}), we adopt our results from the K66 model
for the analysis and discussion following in the later sections.
The K66 reductions also allow easier comparison with characteristic-size
GC data from other galaxies, which we use in the later sections.
As is discussed above, the W75 model becomes most effective for GCs
that happen to have extended, low-surface-brightness envelopes,
which would become noticeable only beyond our radial measurement 
limit of $\simeq 0.5'' \simeq 20$ pc, and even then only for
the most luminous clusters whose envelopes would be detectable
above the sky noise.
Higher-S/N data than we have at present will be needed to trace these outer parts for
most clusters (with the notable exception of the UCD discussed above).

\begin{figure}
\includegraphics[angle=0,width=0.5\textwidth]{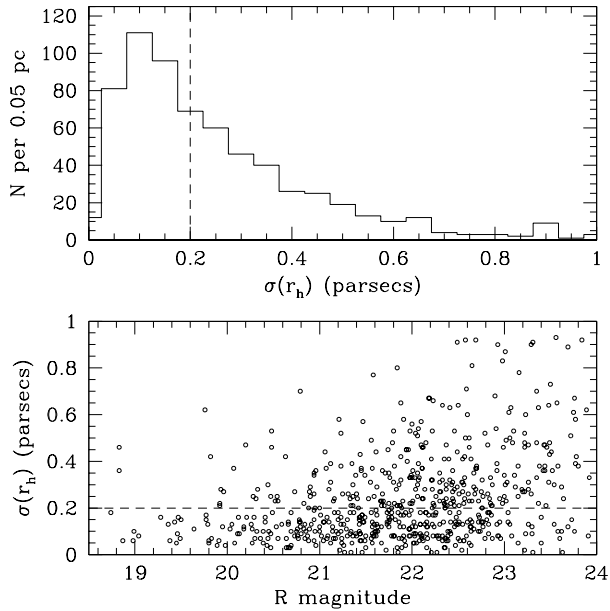}
\caption{\emph{Upper panel:} Histogram of
uncertainty $\sigma(r_h)$ in cluster
effective radius returned by the model fit, in parsec units.
\emph{Lower panel:}  Fitting uncertainty versus cluster brightness.
The median uncertainty of 0.20 pc is shown as the dashed line.
}
\label{errors}
\end{figure}

To obtain a final set of $r_h$ values, we took an unweighted average
of the measurements in all three filters and calculated the uncertainty
of the mean as equal to 
$\sigma(r_h) = \lbrace \sum (r_{hi}-\langle r_h\rangle)^2 / n(n-1) \rbrace ^{1/2}$.
The distribution of these uncertainties is shown in Figure \ref{errors},
along with their dependence on $R$ magnitude.
As expected, the average $\sigma(r_h)$
increases for fainter objects due to lower S/N and
the increased relative effect of background noise.  The \emph{median}
uncertainty over the entire dataset is $\pm 0.20$ pc.  

\begin{figure}
\includegraphics[angle=0,width=0.5\textwidth]{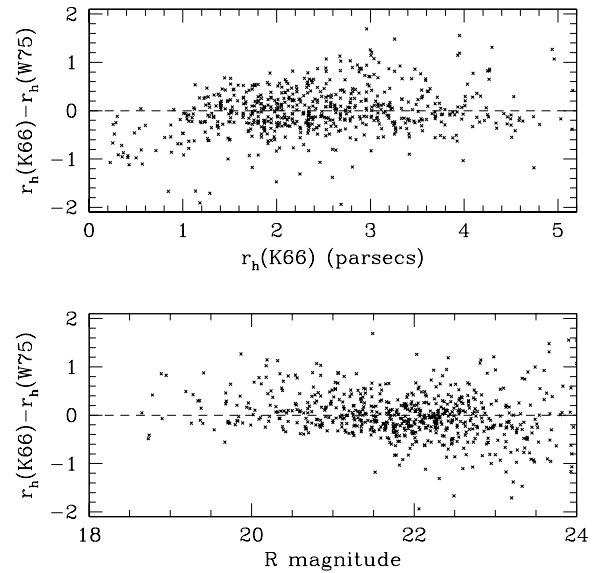}
\caption{\emph{Left panel:} Difference between the cluster sizes
$\Delta r_h = (r_h({\rm King}) - r_h({\rm Wilson})$ plotted versus
size.  Both scales are in pixel units (1 px = $0.05''$).
\emph{Right panel:}  $\Delta r_h$ versus cluster brightness.
}
\label{rhdiff}
\end{figure}

Next, in Figure \ref{rhdiff}, we show the difference in the average cluster
size $\Delta r_h = (r_h({\rm K66}) - r_h({\rm W75}))$ as functions of
cluster size and brightness.  The \emph{median}
$\Delta r_h$ is 0.02 pc, indicating no important systematic difference
between the two models over the entire range of the data.
There is a slight tendency for $\Delta r_h$ 
to vary nonlinearly with either size or brightness.  However,
these trends fall within the internal scatter, and the most obvious
interpretation is simply that at some level,
we reach the irreducible ``floor'' where the fundamentally different
assumptions built into the different models can lead to slight
differences in the best-fit structural parameters.  These parameters are
unavoidably model-defined, and at this level the question about which
solution represents the ``true'' cluster size becomes moot.  In general, we take this graph
as useful primarily for estimating the internal uncertainty of the fit
because of the model assumptions alone.  
This fitting uncertainty per cluster (which we adopt as the rms scatter
in the graph) is then $\pm 0.21$ parsec.

\begin{figure}
\includegraphics[angle=0,width=0.5\textwidth]{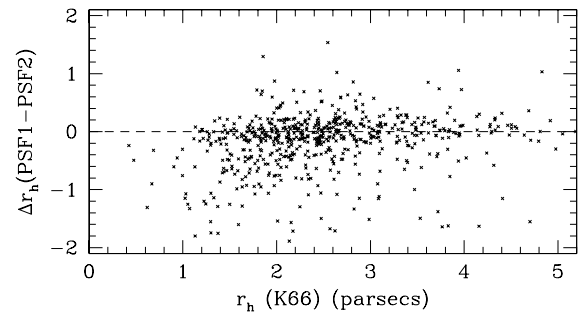}
\caption{Difference between the measured cluster sizes for
two different assumptions about the PSF star.  One run (denoted
here PSF1) uses
the nearest moderately bright star to each cluster as the fiducial
PSF for that cluster; the second run (PSF2) uses the second-nearest star.
}
\label{psfcomp}
\end{figure}

A final internal check of our reductions is to gauge the uncertainty
in size measurement due to the PSF itself.  Our ``nearest star'' approach to defining the
PSF for each target cluster ensures that the results will not be
affected by large-scale trends in PSF size across the mosaic, but
any one PSF star will have lower S/N than the average of many of them
across the field and thus slightly higher random uncertainty.
To evaluate this level of uncertainty, we ran the model fits
a second time, now using the \emph{second-nearest} star for each object.
Direct comparison of the two reductions is shown in Figure \ref{psfcomp}.
Again, no systematic difference appears bigger than 0.05 pc, but
the rms scatter is $\pm 0.30$ pc.

In summary, all three of the possible sources of fitting uncertainties
(filter-to-filter consistency, choice of profile model, choice of PSF)
contribute to the net uncertainty to about the same level.  Adding the
three in quadrature,
we then estimate that a global-average uncertainty in the cluster
sizes, due strictly to the internal measurement process, is $\pm 0.41$ parsec. 
As will be seen below, this is equivalent
to about 16 percent of the median cluster size.

It is worth noting here that the mean uncertainty of $\pm 0.4$ pc also allows
us to resolve an important feature of the GC size distribution, namely
its lower limit.  As will be seen in Section 3, the SDF starts increasing
sharply near $r_h = 1$ pc; clusters smaller than this are 
almost nonexistent in M104 or in other galaxies.  For the Virgo galaxies
at $d=16.5$ Mpc, almost twice as far away as M104, this lower edge to the SDF is not 
well determined \citep{jor05} except with extremely high$-S/N$ data
\citep{madrid09}.  For the gE galaxies at $d > 30$ Mpc studied by 
\citet{harris09}, the edge falls below the limits of HST resolution and
only the upper half of the SDF can be clearly measured.  The relevance of this
feature to understanding the formation of GCs and the origin of the SDF
will be discussed in Section 4 below.

\subsection{Comparisons with Paper I}

Lastly, we compare the results from our
new model profile fits with those done in Paper I, which used
the ISHAPE code \citep{lar99} with the analytic K62 model profile.
For this test, we carried out a separate run of the \citet{mcl08}
code used above, but now using the K62 model in order to rule out
any differences due only to the adopted model and focus on the two
codes themselves.

\begin{figure}
\includegraphics[angle=0,width=0.5\textwidth]{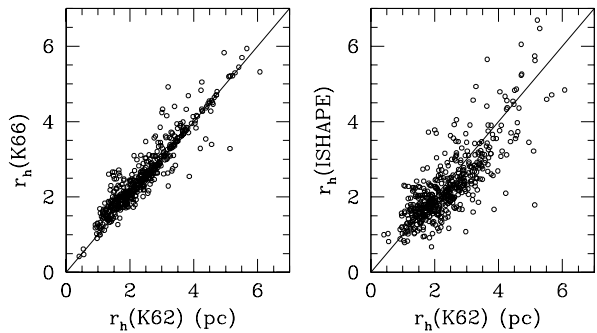}
\caption{\emph{Left panel:}  Cluster effective radius $r_h$ from the King (1966)
profile fits, plotted versus our average $r_h$ values determined
from the simpler King (1962) analytic model.
\emph{Right panel}:  Cluster effective radius as determined from
ISHAPE in Paper I, plotted versus the King (1962) model fits
determined in the present paper.
A 1:1 correlation line is drawn in for each graph.
}
\label{ishape}
\end{figure}

In the first panel of Figure \ref{ishape}, we show the direct
comparison of this run with our K66 fits.  
Though there is overall close agreement, the median difference
is $0.19$ pc in the sense that the K66 models tend to measure
the clusters slightly (8 percent) larger.  The rms scatter is
$\pm 0.32$ pc, similar to the internal comparisons already described above.

The second panel of Figure \ref{ishape} shows the correlation of
our new K62 fits with those from ISHAPE and Paper I.
For \emph{small} objects ($r_h \la 2.5$ pc, or effective radii
less than about 1 pixel on the images) the two methods
agree systematically quite well.  
However, for larger objects ($\ga 2.5$ pc) there is an offset
between the two codes in the sense that ISHAPE measures them smaller
by about 0.3 pc than does the McLaughlin et al. code.
One difference between the two runs is that the ISHAPE reductions
assumed a fixed central concentration of $c = {\rm log} (r_t/r_c) = 1.5$ for all
clusters, whereas the McLaughlin code solves for $c$ (equivalently, 
the central potential $W_0$) as a free parameter.  However, this
should mainly introduce cluster-to-cluster scatter since $c  \simeq 1.5$
is a reasonable average for real GCs.  A possible cause for a systematic
offset may be 
(as is described in Paper I) that the ISHAPE sizes in the three filters
were all normalized to the $V-$band data and then averaged,
but it is not clear why a magnitude-dependent offset should appear.
These two factors aside, the remaining differences are presumably due
to the details of the two codes, and we use this
test only as a rough consistency check.  We conclude that at the 10-20\% level,
both codes and all three models return similar
intrinsic cluster sizes.  As will be discussed in the next section,
most trends found in Paper I, such as cluster size versus magnitude or 
galactocentric radius, fall into the same patterns here.

\section{The Size Distribution}

\begin{figure}
\includegraphics[angle=0,width=0.5\textwidth]{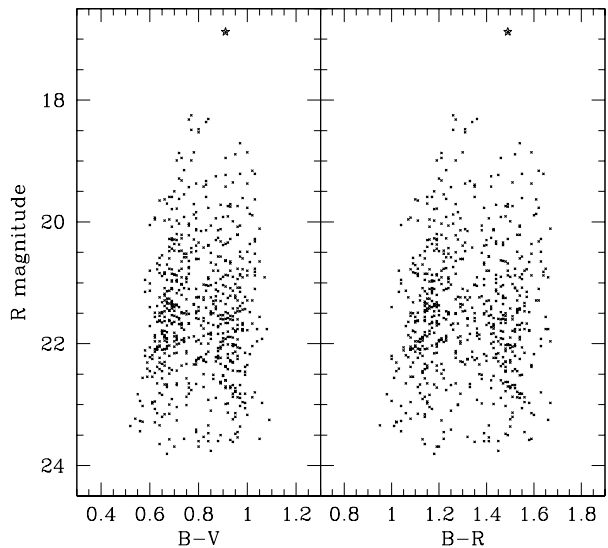}
\caption{Magnitude $R$ versus color index $(B-V)$ or $(B-R)$ for
659 globular clusters in M104.  The single known UCD is shown at
top as the large starred symbol \citep[see][]{hau09}.  All $BVR$ magnitudes are
those measured through 5-px aperture individually corrected
to 20-px radius for cluster size as described in the text
and Figure \ref{apcorr}.  
}
\label{cmdraw}
\end{figure}

The color/magnitude diagrams in $(R,B-V)$ and $(R,B-R)$ are shown
in Figure \ref{cmdraw}.  The classic bimodal division into the
blue, metal-poor and red, metal-rich sequences is easily
visible, with a (more or less arbitrary) division at $(B-R) \simeq 1.30$.
A closer analysis of these bimodal sequences is presented 
in Section 5 below;
first, we take a closer look at the distribution of GC sizes
as a whole, and their correlation
with external properties including luminosity, metallicity,
and projected galactocentric distance.

\subsection{The Overall Scale Size Distribution}

\begin{figure}
\includegraphics[angle=0,width=0.5\textwidth]{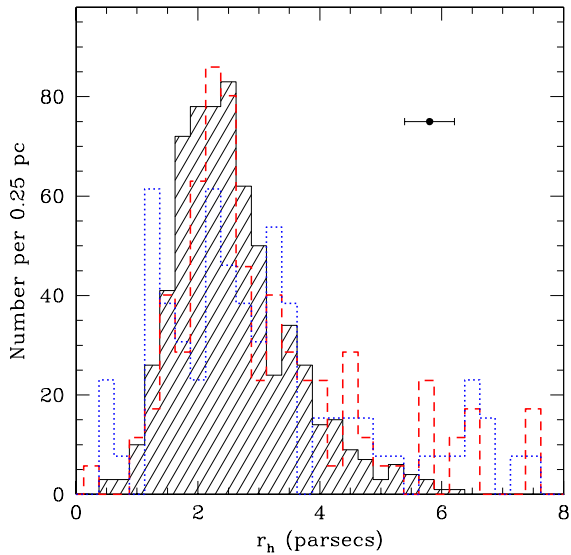}
\caption{Histograms of globular cluster scale sizes for three
galaxies.  Our present results for M104 are shown
as the \emph{shaded histogram}; the Milky Way GCs are
in shown as the \emph{dashed red histogram}, and the compilation
of old GCs from nearby dwarf galaxies \citep{georgiev08} as
the \emph{dotted blue histogram}.  The Milky Way and dwarf
samples have been normalized to the same total population
as in M104.  The errorbar at upper right is the mean internal
uncertainty of $\pm0.41$ pc (see text) for our measurements.
}
\label{rh_histo}
\end{figure}

The overall distribution of the GC scale sizes in our dataset
is shown in histogram form in Figure \ref{rh_histo}.  The
distribution is modestly skewed to larger $r_h$, with a median
at $2.44 \pm 0.04$ pc and an intrinsic dispersion of $\pm 0.85$ pc.

To calculate the dispersion for this and for other
distributions used in this paper,
we adopt here the Median Absolute Deviation (MAD), a robust estimator
of the intrinsic scatter of a data sample.  It is useful in cases where the distribution is 
asymmetric and even for cases where the conventional standard
deviation may be formally undefined \citep[e.g.][]{hoag83}.  For a dataset
$\lbrace x_i \rbrace$ with median $\tilde x$ the MAD is defined as
\begin{equation}
MAD = {\rm median}\lbrace \vert x_i -  \tilde x \vert \rbrace 
\end{equation}
and the sample dispersion is then estimated as
\begin{equation}
\sigma \, = \, 1.483 \times MAD .
\end{equation}
For a Gaussian distribution, this formula
exactly gives the usual standard deviation.

A basic point of immediate interest is to compare the GC
size distribution with the one for the ``baseline'' Milky Way
system.  However, a minor spatial bias must be kept in mind.
The Sombrero GCs in our list do not cover its entire halo,
whereas (in the Milky Way at least) there is a well known
trend for $\langle r_h \rangle$ to increase systematically with $R_{gc}$
\citep{vdb91}.
The GCs in our data have projected galactocentric distances
ranging from $R_{gc} \simeq 0.9$ kpc out to 15 kpc, with reasonably complete
radial coverage to 8.5 kpc.  To extract a Milky Way sample
that will more closely mimic the M104 data, we take the
114 known Milky Way clusters with measured half-light radii 
obtained from King-model fits \citep{har96} and with
\emph{projected} Galactocentric distances $R_{gc} = (Y^2 + Z^2)^{1/2} < 12$ kpc.
Here $(Y,Z)$ are the distance components projected on the sky parallel to and perpendicular
to the Galactic plane.\footnote{The third component $X$ is directed
along the axis from the Sun to the Galactic center, and contains
most of the random errors in the distance measurements to the
individual clusters \citep[see][]{racine89}.  By projecting them onto the $YZ$ plane
we therefore get the closest to seeing the system as if it were an
external galaxy free of position-measurement bias.}  In Figure \ref{rh_histo} the distribution for
the 114 selected Milky Way clusters is shown as the dashed red histogram.
The median is at $r_h(MW) = 2.61 \pm0.10$ pc, and to first order, the
two distributions are strikingly similar.  The intrinsic spread
of cluster sizes around the peak is characteristically $\simeq \pm 1$ pc rms,
and very few clusters exist with physical scale sizes smaller than about 1.0 pc.
The slightly broader width
of the central peak for M104 is likely to be due in part
to the instrumental broadening of $\pm 0.4$ pc. (However, it should also
be noted that the Milky Way data come from a somewhat
heterogeneous collection of starcount and surface brightness data
obtained in many programs for clusters at very different distances
from the Sun).

Another dataset making an interesting comparison is the recently
measured set of old GCs in several dwarf galaxies, from
\citet{georgiev08}.  On average these small galaxies are at 
similar distances to M104, and their profiles were measured
from HST/ACS images with ISHAPE+K62 profile fits.  \citet{georgiev08}
give data for 83 objects regarded to be classically ``old'', luminous
GCs.  Of these, 42 clusters come from dIrr galaxies, 31 from dSph and dE systems,
and 12 from Sm galaxies.  The size distribution of these is plotted in Figure \ref{rh_histo}
as the blue dotted histogram.  The median of this dwarf-galaxy
sample is at $r_h = 2.87 \pm 0.19$ pc.

Although the medians or the peak points of these three samples
are not strongly different, a potentially more important test is the total
shape of the whole distribution.  A standard Kolmogorov-Smirnov two-sample
test shows that the M104 and Milky Way GCs are significantly
different at the 97\% level, whereas the M104 and dwarf samples
are different at more than 99\% confidence (the Milky Way and dwarf
samples are \emph{not} significantly different from 
each other).\footnote{The
difference between the M104 and dwarf samples would be even stronger if,
as is hinted by Figure \ref{ishape}, the ISHAPE fits need to be corrected
to slightly larger $r_h$ to put them onto the same internal scale as the
present code.}
The key difference among these histograms is the relative
number of ``extended'' clusters with $r_h \ga 5$ pc, i.e. the ones more
than twice the median size.  Our initial selection of 
GC candidates in M104 did not rely on
object scale size except for rejection of \emph{small}, starlike objects
(see Paper I), so the sample should not be biased
against GCs in the range $r_h \sim 5-10$ pc or even larger.  Nevertheless, even
a few more objects added to the high-end tail of the distribution
would noticeably reduce the statistical difference between the Milky Way and
M104, so for the present, we regard these comparisons as only indicative.
Perhaps a more important conclusion is that if we use \emph{only the
clusters smaller than 5 pc}, there are no significant differences
among these three samples.

Extended clusters would clearly find it easier
to survive in the gentler tidal environment of smaller galaxies,
or in the outer-halo regions of large galaxies.
If we take the comparisons in Figure \ref{ishape} at face value,
M104 -- a massive, bulge-dominated Sa galaxy -- 
has very few such extended GCs compared with the other two samples.  
Possible interpretations that immediately
suggest themselves are \emph{either} that M104 
did not acquire most of its globular cluster population by accretion of
small satellites; \emph{or} that any extended clusters that might have
been accreted this way have already been tidally destroyed.
A survey of the outer parts of M104's halo would provide
much clearer evidence to discuss this argument further.

\subsection{Correlations with Metallicity}

Previous surveys of GCs in nearby galaxies have shown
that the blue, metal-poor clusters are consistently larger
on average than the red, metal-richer ones
\citep[][]{kun98,kun99,larsen01,jor05,spitler2006,gom07,harris09}.
These differences are at the level of only 15-20\%, but
in the biggest samples \citep[e.g.][]{larsen01,jor05,harris09}
they are highly significant in a statistical sense, and
they are found at all galactocentric distances.
\citet{harris09} determines a mean size difference of $(17 \pm 2)$\%
(blue \emph{minus} red) from a sample of several thousand clusters in 
six supergiant ellipticals (Brightest Cluster Galaxies or BCGs).

\begin{figure}
\includegraphics[angle=0,width=0.5\textwidth]{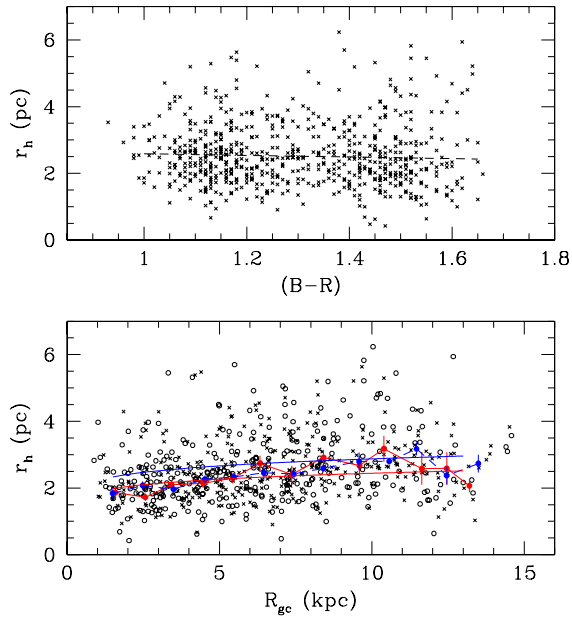}
\caption{\emph{Upper panel:} Scale sizes of M104 clusters
versus $(B-R)$ color.  A linear fit to the points is shown as
the dashed line.
\emph{Lower panel:}  Scale sizes for the clusters as a function
of projected galactocentric distance $R_{gc}$.  Blue-sequence clusters
($B-R \leq 1.3$) are plotted as crosses, red-sequence clusters 
($B-R > 1.3$) as open circles.  Large filled circles with errorbars
denote the median sizes in 1-kpc bins, separately for the blue
clusters (dashed blue line) and red clusters (solid red line).
The two smooth curves (blue, red) show the scaling laws
$r_h \sim R_{gc}^{0.11}$ that have been found for globular
clusters in supergiant ellipticals \citep{harris09}.
}
\label{rh_rgc}
\end{figure}

Our observed correlation is shown in the upper panel of
Figure \ref{rh_rgc}.  The red clusters do lie at lower
sizes on average, but an unweighted linear fit to the complete dataset
gives a slope $\Delta r_h / \Delta(B-R) = -0.245 \pm 0.191$,
which is not highly significant.  (See also Paper I for a
similar diagram.)  For all blue clusters
(defined here as those with $(B-R) \leq 1.30$), the median
size is $r_h(blue) = 2.47 \pm 0.04$ pc, while the redder
clusters ($B-R) > 1.30$) have a median $r_h(red) = 2.33 \pm 0.05$ pc.
The metal-poor GCs are thus 6\% larger, and the difference
$\Delta r_h = 0.14 \pm 0.064$ pc is significant at the 
$2.2 \sigma$ level, a tantalizing but not conclusive offset.
In Paper I, the difference between median sizes was found
to be $\Delta r_h = 0.25 \pm 0.06$ pc, similar to the present
value.  Thus we find the same 
effect seen in other systems, but at a smaller amplitude.

To date, the strongest claims for metallicity-related differences in size
have relied on the large GC samples from
\emph{elliptical} galaxies.  The samples from
GCs in \emph{disk} galaxies are much smaller and inevitably have
lower statistical significance. Notably, however, \citet{can07}
and \citet{deg07} find that the mean size difference for the GCs
in the disk galaxies NGC 5866 and 1533 is at the level
of $\sim 0.1$ pc or less, quite similar to our results for M104.

Correlation of mean size with location in the halo has
been proposed as an important factor in deciding what the
physical cause for the metallicity/size offset actually is
\citep{lar03,jor04}.  If the size difference versus metallicity persists at
all galactocentric distances, then it is less likely to be
due to a geometric projection effect (the metal-richer clusters 
are usually found to have a more centrally concentrated spatial distribution
than the metal-poor ones, thus will be more subject to stronger
tidal stripping.  This effect would be much stronger on
the inner-halo clusters, and not as important for the clusters
of both types that are found in the outer halo).  
In M104, as shown in Paper I, both
red and blue GC subsystems accurately follow a $r^{1/4}-$law
profile, 
$$ {\rm log} \sigma_{GC} \, = \, \alpha R_{gc}^{1/4} \, + \, \beta \, .$$
For the red GCs, using our present data over the radial
region $R_{gc} < 3'$ where we have complete azimuthal coverage, we find
$\alpha = -1.937 \pm 0.046, \beta = 3.191 \pm 0.053$; while
for the blue GCs,
$\alpha = -1.335 \pm 0.179, \beta = 2.541 \pm 0.205$. 
Thus the blue, more metal-poor subsystem is significantly
more extended.  According to the geometric projection hypothesis,
we would then expect a significant mean size difference between
blue and red in the inner regions but less so in the outer halo.

This version of the size distribution
is shown in the lower panel of
Figure \ref{rh_rgc}, separately for the blue and red clusters.
Median values for $r_h$ are plotted as the connected large points
with errorbars, in 1-kpc bins.  
The GCs of both metallicity groups have scale sizes that increase
gradually but consistently throughout the halo of the galaxy.
The slopes of both trends are similar, so if we combine all clusters to
gain statistical weight, we find a rate of increase of mean
cluster size $r_h \sim R_{gc}^{0.17 \pm 0.02}$.
The effective radius of the spheroid light is $R_{eff} = 1.2' = 3.14$ kpc,
so the total radial range of our data reaches to 
an outer limit of about $4.5 R_{eff}$.  Not only does the radial
increase affect all clusters, but the difference between red
and blue remains similar (and small) at all radii.  
This large-scale trend may therefore provide evidence against
the geometric-projection effect \citep{lar03} as being the sole
explanation for the size difference in this galaxy.  However, a more 
detailed deprojection model will be needed to test this
conclusion more quantitatively \citep{lar03}.

The only other data in the literature that cover similarly large
ranges in $R_{gc}$ are for NGC 5128 \citep{gom07}, which
reach even further to $\simeq 8 R_{eff}$ but cover a smaller sample; 
and for the six supergiant ellipticals studied by \citet{harris09},
which extend to distances $\simeq 4.5 R_{eff}$. 
Both of these other surveys, however, indicate the same steady 
increase in scale size with $R_{gc}$ for clusters of both types.
So do the smaller samples in the two disk galaxies mentioned
above \citep{can07,deg07}.
\citet{harris09} derives a simple power-law scaling for gE's of
$r_h(med) = r_{h0} (R_{gc}/R_{eff})^{0.11}$, where
the zeropoints are $r_{h0} = 2.53$ pc (blue) and 2.15 pc (red).
These functions are shown in Figure \ref{rh_rgc}(b).  Clearly,
they strongly resemble our current data, bracketing the median
points and indicating that the M104 GCs follow a very
similar trend.  

The total evidence suggests that the dependence of
GC scale size on metallicity is at least partly intrinsic to
the clusters, and thus due to a more \emph{local} cause having
to do either with their formation or later internal evolution.
\cite{jor04} has proposed that it is the result of stellar-evolution
timescales that depend on metallicity, coupled with many internal
relaxation times of dynamical evolution that would make the
metal-rich clusters appear smaller.  

Another possibility is simply that the metal-richer
clusters benefitted from more rapid cooling and contraction while
they were still gaseous protoclusters and had not yet formed
most of their stars \citep{harris09}.  Still another possibility
is that all clusters started out with similar sizes during their
early protocluster stage, but the star formation efficiency (SFE) was
a bit higher for more metal-rich gas, allowing the cluster to
expand less during the gas expulsion phase (see Section 4 below). 
These alternatives
will be difficult to compare quantitatively, but detailed models
would be of great interest.

\subsection{Size versus Luminosity}

\begin{figure}
\includegraphics[angle=0,width=0.5\textwidth]{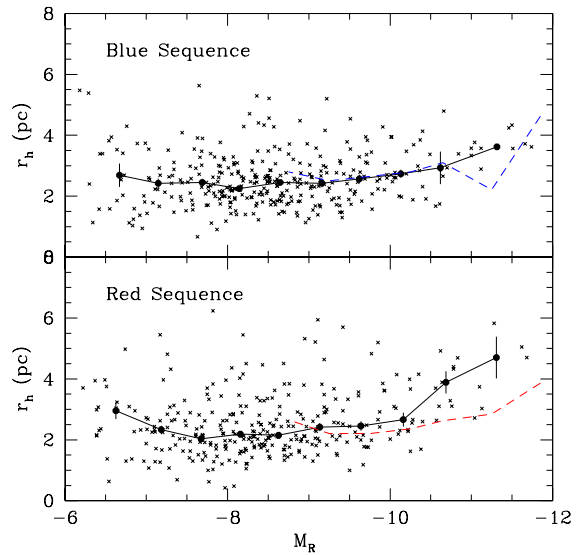}
\caption{\emph{Upper panel:} Sizes of the blue-sequence clusters
($(B-R) < 1.3$) as a function of luminosity $M_R$.  The connected solid dots with
errorbars show the median $r_h$ in half-magnitude bins (see text).  
\emph{Lower panel:}  Sizes for the red-sequence clusters
($(B-R) > 1.3$).  In both panels, the mean relations for GCs in six
supergiant ellipticals \citep{harris09} are superimposed as
the dashed lines.
}
\label{redblue}
\end{figure}

The correlation of cluster scale size with 
luminosity is the observational version of the mass/radius relation.  
This form is displayed in Figure \ref{redblue}.  Here, the individual clusters
are plotted along with the median $r_h$ in half-magnitude bins
of luminosity $M_R = R - (m-M)_R$ where our adopted apparent distance modulus is
$(m-M)_R = 29.90$.  Over the range $-7 \ga M_R \ga -10$ (corresponding
approximately to the luminosity range $3 \times 10^4$ to $6 \times 10^5 L_{\odot}$)
the median size remains nearly constant, increasing more rapidly for
luminosities $M_R \la -10$.    The pattern for the GCs in the supergiant
ellipticals from \citet{harris09}, shown as the dashed lines in Figure \ref{redblue},
is the same, although these six galaxies are $\sim 5 \times$ more distant
than M104 and thus the GCs could not be traced to similarly faint luminosities.
Here, the $(I, B-I)$ photometry used for the giant ellipticals has been
converted approximately to $M_R$ with the assumption $R-I \simeq 0.5$.

\begin{figure}
\includegraphics[angle=0,width=0.5\textwidth]{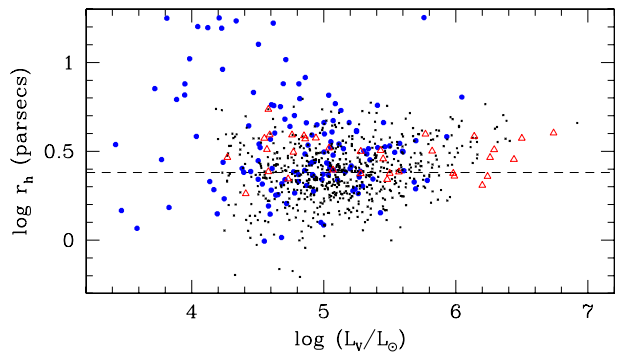}
\caption{Scale sizes of globular clusters in three massive
disk galaxies:  M104 (small crosses), the Milky Way (large
filled circles), and M31 (triangles).  The median size
of 2.4 pc is drawn in as the dashed line. Note the systematic
rise in median $r_h$ for luminosities above log $(L_V/L_{\odot})$ $\ga 5.5$,
as well as the scatter to high $r_h$ at low luminosities.
}
\label{rhmag}
\end{figure}

In Figure \ref{rhmag}, the data for M104 are plotted in comparison with
two other large disk galaxies, the Milky Way and M31 \citep[with data from][]{bar07}.
This form of the graph shows more clearly the trend for the most luminous GCS
($L \ga 3 \times 10^5 L_{\odot}$),
extending up to the very most luminous GCs known near $10^7 L_{\odot}$.
A more extensive discussion of the possible link between
massive GCs and the still more massive UCDs is made by 
\citet{has05} and has been developed further
in, for example, \citet{kis06,evs08,for08}.  In general, UCDs have scale sizes
of $\simeq 10$ pc and above, thus sit a bit higher on the $(L,r_h)$
relation than do the GCs.  The single UCD known in the M104 field is
more than one magnitude brighter (Figure \ref{cmdraw}) than the
top end of either the blue or red GC sequences.  

A key factor 
distinguishing a massive GC from a UCD or nuclear cluster may be 
the higher mass-to-light ratio for UCDs \citep{mie08,baum08}.
Multiple cluster mergers are another route to forming supermassive GCs
and UCD-like objects and can explain the observed upturn in the $L$ vs. $r_h$ 
correlation \citep[e.g.][]{kis06}; but it is less clear whether such mergers
would also produce objects with increased $(M/L)$.
It remains possible that at least some UCDs may simply be very massive GCs.
This latter view is supported by \citet{mur09}, who develops a model for
the size/mass/luminosity relation using the idea that very massive
protoclusters ($\ga 10^6 M_{\odot}$) will be optically thick to their
far-IR radiation, making radiation pressure important for energy balance.  The resulting
increase in Jeans mass with cluster mass will yield a scaling of $M/L$
with mass that matches the trend for UCDs and the most massive GCs reasonably well.

\subsection{Fundamental-Plane Quantities}

Recent work has established the existence of a surprisingly narrow
``fundamental plane'' (FP) of structural quantities for globular clusters
\citep[e.g.][]{djo95,mcl00,mcl05}.  The so-called $\kappa-$space 
of three orthonormal quantities $(\kappa_1, \kappa_2, \kappa_3)$ 
is conventionally constructed from the central velocity dispersion $\sigma_{p0}$,
effective radius $r_h$, and surface mass density $\Sigma_h$.
For GCs in galaxies such as this one beyond the Local Group, direct velocity
dispersion measurements (hence mass) are rare by comparison with 
photometry, and in any case require knowing the cluster core radii $r_c$ 
to convert the integrated velocity dispersion to its core value.
The core radii in turn have to be estimated from $r_h$ and the model-fitted
central concentrations $c$, which are quite uncertain for objects such as these
where $r_c$ is far smaller than the image resolution.

\begin{figure}
\includegraphics[angle=0,width=0.5\textwidth]{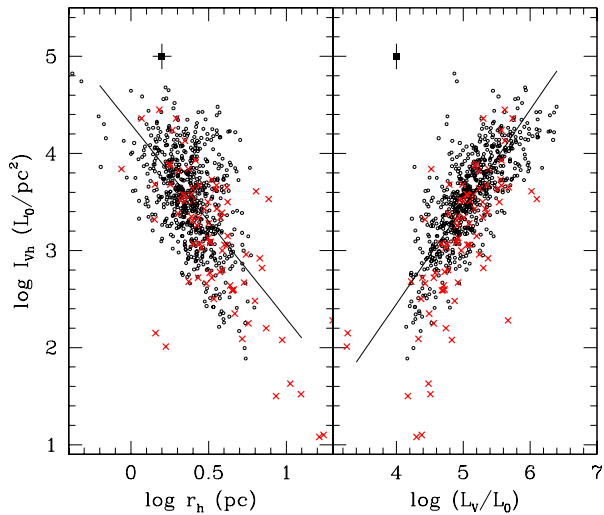}
\caption{Surface intensity in $V$ over the half-light radius,
plotted as a function of cluster scale size (left panel) and
total luminosity (right panel).  Small dots give the data
for the M104 clusters, while the large red crosses give similar
data for the Milky Way clusters \citep{mcl05}.  The large square
at upper left in each panel with errorbars shows the mean 
internal uncertainty in each datapoint.  The straight lines in
each graph show the nominal proportionality scalings $I_h \sim r_h^{-2}$
and $I_h \sim L$.
}
\label{iv}
\end{figure}

Although a full discussion of the FP in all its various guises is 
therefore not possible, more limited
versions can still be constructed.  A useful quantity representing the
cluster density is the surface intensity $I_h = L_{tot}/(2\pi r_h^2)$
integrated over the half-light radius.  This quantity is well enough determined
to allow us to compare the M104 GC population directly with the baseline
Milky Way system \citep[see also][for similar data in M31 and other Local Group members]{bar07}.
Our measurements give internal uncertainties of $\pm0.1$ mag or better
in luminosity and $\pm0.4$ pc in radius, making (log $I_h$) uncertain to $\pm 0.13$.
In Figure \ref{iv}, $I_{Vh}$  is plotted
against both cluster size and total luminosity.  The analogous Milky Way
data are taken from \citet{mcl05}.  The obvious trends that $I_h$ is brighter
for more luminous or more compact clusters are evident from the graphs,
but the main conclusions to draw are that the M104 GCs match up well
with both the mean positions of the Milky Way clusters and the intrinsic
scatter around the fiducial scaling lines (shown in the graph).
The clusters at lowest luminosities and surface brightnesses do, however,
tend to have systematically larger scale sizes, dropping them below
the fiducial scaling lines in both graphs.  The papers listed 
above give more detailed discussions of these relations.

An alternate and conceptually simple formulation of the FP is in terms
of the binding energy of the cluster, $E_b = f(c) G M^2 / R$ \citep{mcl00},
where $f(c)$ represents the details of the internal mass distribution
for a given cluster and $R$ is some characteristic radius.
As is shown in \citet{mcl00} and \citet{bar07}, for normal globular clusters
the binding energy varies almost exactly as $E_b \sim M^2$ with remarkably little relative
scatter, a scaling law that is basically quite different from
the $E_b \sim M^{1.5}$ rule characterizing other structures such as giant
molecular clouds or E galaxies.  This form of the FP has been used for 
globular clusters within Local Group galaxies 
including the Milky Way, M31, and the Magellanic Clouds \citep{bar07}.
As originally defined by McLaughlin, the calculation
of $E_b$ requires fairly precise knowledge of the King core radius and central
concentration, which we do not have for M104 and other galaxies at large
distances.  However, \citet{mcl00} also shows that 
it can be recast in terms of the effective (half-light) radius,
which is much more accurately known.  The
major advantage of defining $E_b$ this way is that it allows us to
extend the FP discussion to the huge cluster populations in the giant
galaxies that lie in the near-field region beyond the Local Group.

An appropriate combination of the equations in McLaughlin's paper gives
\begin{equation}
E_b \, = \, G ({4 \pi \over 9})^2 { {\cal R} {\cal E} \over {\cal L}^2 } {M^2 \over r_h} \, 
\end{equation}
where ${\cal R}, {\cal E}, {\cal L}$ are all dimensionless functions of $c$
that can be calculated from the prescriptions in \citet{mcl00}.
Happily, the ratio (${\cal R} {\cal E} / {\cal L}^2$) 
is nearly constant \citep[see also][]{mcl00}, equalling $0.17 \pm 0.02$ over
the range of King $c-$values appropriate for normal GCs, so by using the 
half-light radius we do not need to know the central concentration precisely.  
A potentially more important \emph{caveat} is that strictly speaking, knowing $M$
requires also knowing the cluster mass-to-light ratios.  In turn, measurement of $M/L$ 
independently of the photometry or colors requires direct
measurement of the internal velocity dispersions, which are not yet available for M104
with the exception of the bright UCD (see below).
For the present purposes, we simply assume $(M/L)_V = 2.0$ and then calculate $E_b$
for all the M104 clusters in our list.  This fiducial $(M/L)$ is a compromise
choice drawn from the recent literature for dynamically measured masses of 
GCs in the Milky Way, M31, and NGC 5128 \citep{mcl05,rej07,str09,baum09}.
These measurements fall in the typical range $(M/L)_V \simeq 1 - 3$,
probably with real cluster-to-cluster differences depending on metallicity, luminosity,
or galactocentric distance.

\begin{figure}
\includegraphics[angle=0,width=0.5\textwidth]{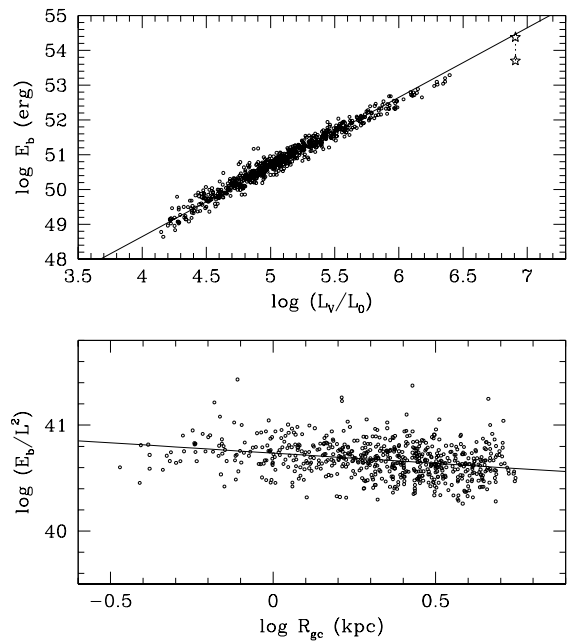}
\caption{\emph{Upper panel:} Binding energy $E_b$ measured in ergs,
plotted against cluster $V-$band luminosity in Solar units $L_V/L_{\odot}$.
The straight line is an exact scaling law $E_b \sim L^2$.  The two large
starred points at upper left show the position of the UCD under two
different assumptions about its M/L ratio (see text).
\emph{Lower panel:}  Normalized binding energy $E_b/L^2$ versus projected
galactocentric distance.  The fitted straight line has a slope 
of $-0.19$ (see text).
}
\label{eb}
\end{figure}

In Figure \ref{eb}, the resulting correlation of binding energy with cluster luminosity is
shown.  A direct least-squares solution gives log $E_b/{\rm ergs} = (41.11 \pm 0.07) + (1.92 \pm 0.02) {\rm log} L$,
with a scatter of $\pm 0.16$ dex.  For the Milky Way GCs
McLaughlin found slopes in the range $1.8 - 2.2$ under various assumptions. 
In summary, we concur that the scaling rule $E_b \sim L^2$ provides an excellent first-order description
of the data.  The scatter around the line in Figure \ref{eb}a is artificially small (only
half as large as for the Milky Way sample) because it does not account for cluster-to-cluster
differences in the $M/L$ ratio that must be present (and since $E_b$ varies as $L^2$ it is
moderately sensitive to changes in $M/L$).  

The UCD in the M104 field is of special interest as a possible ``connector'' to the GC sequence.
Its position is shown in Figure \ref{eb} at upper right.
The lower of the two connected points is where we would have located it if we had assumed
$(M/L)_V = 2.0$ as we did for all the GCs.  The upper point uses its actual
value of $(M/L)_V = 4.36$ as directly measured from its internal velocity dispersion \citep{hau09}.  
Remarkably, its true position on the graph extends the same $E_b$ curve defined by the 
GCs accurately upward by another order of magnitude beyond the top of the GC sequence.
This result, in addition to the other characteristics of the UCD measured by \citet{hau09},
is consistent with the interpretation that this luminous, compact system is a 
massive GC.

The binding energy also provides a sensitive confirmation test of changes in
cluster structure with environment.  
\citet{mcl00} found that the residuals from $E_b$ vs.~$L$
were a significant function of galactocentric distance in the Milky Way (see his Fig.~10),  
in the sense that clusters further out in the halo are less
tightly bound because of their systematically larger radii.  
Using the results from Figure \ref{rh_rgc}  and Section 3.2, 
we can predict that for our M104 data, $(E_B/L^2)$ should vary as $R_{gc}^{(-0.17\pm 0.02)}$.
The consistency test is shown in Figure \ref{eb}b, where we find
a net downward trend with a fitted slope of $(-0.19 \pm 0.02)$ in good agreement
with the prediction.  For the Milky Way,
McLaughlin found a slope of $-0.4 \pm 0.1$ for the trend of $E_b/L^2$ versus
\emph{three-dimensional} Galactocentric distance, which after
projection to two dimensions will decrease the slope to a level closer to
our result.

\begin{figure}
\includegraphics[angle=0,width=0.5\textwidth]{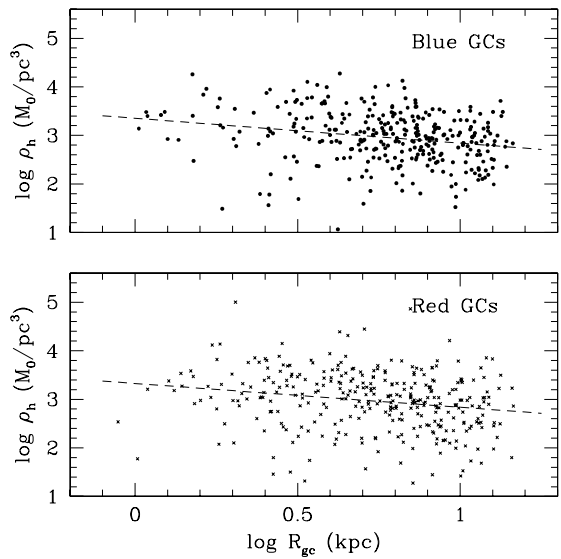}
\caption{Calculated mean density $\rho_h$ plotted as a function
of galactocentric distance.  The metal-poor and metal-rich clusters
are plotted in separate panels, with the mean relationships derived
in the text shown as dashed lines.  
}
\label{density}
\end{figure}

Yet another way to represent the trend of cluster structure with galactocentric
distance is as used recently by \cite{mcl08b} and \citet{chandar2007}.
A characteristic mean internal mass density for each cluster can be
calculated from $\rho_h =  (M/2) \times 3/(4\pi R_h^3)$ where $R_h \simeq (4/3) r_h$
is the three-dimensional half-mass radius.  As above, we use $M/L = 2$ to
transfer from luminosity to mass.  The results plotted separately for
the blue (metal-poor) and red (metal-rich) clusters are shown in Figure \ref{density}.
The best-fit lines through each set of points are
${\rm log} \rho_h (M_{\odot} pc^{-3}) = (3.35 \pm 0.10) - (0.51 \pm 0.12) {\rm log} R_{gc}$ (blue)
and ${\rm log} \rho_h = (3.33 \pm 0.10) - (0.49 \pm 0.13) {\rm log} R_{gc}$ (red).
The scatter around both relations is $\pm 0.55$ in log $\rho_h$.
If mean GC mass does not vary with location in the halo, as the M104 data show, then the
scaling $r_h \sim R_{gc}^{0.17}$ (Section 3.2) would predict
$\rho_h \sim R_{gc}^{-0.51}$, which matches what is found in the density plot.
As is discussed by \citet{chandar2007} and \citet{mcl08b},
the large cluster-to-cluster scatter in density at all galactocentric
distances allows the mean cluster mass to be nearly 
independent of $R_{gc}$ in the presence
of density-dependent dynamical evolution times.

\section{Origin of the Size Distribution}

Aside from minor trends with metallicity, galactocentric distance, and 
mass, the characteristic scale sizes and their distribution function (the SDF) 
in all types of galaxies are remarkably similar over an impressive
range of environments.  Early indications of the near-universal mean or median
GC sizes on observational grounds
were developed a decade ago by \citet{kun01} from HST/WFPC2 measurements
of GC sizes in 28 elliptical galaxies, and by \citet{larsen01}
for a larger sample of GCs in 17 ellipticals.  A major step forward was taken
with the work of \citet{jor05}, who used their database
of $\sim 10^4$ GCs in dozens of Virgo galaxies to strongly reinforce
the view that the full \emph{shape} of the SDF, as well as its mean or median,
is a near-universal characteristic of GCs in galaxies generally.
Furthermore, the SDF has also been found to be similar over a wide range
of cluster \emph{age} \citep[e.g.][]{bar06,sce07}. Its key features are a typical median size
$r_h \simeq 2-4$ pc, a rather sharply defined cutoff below 1 pc, and an
asymmetric tail extending to larger radii.
The near-universality of this distribution across all types of galaxy
environments 
suggests to us that the conditions local to the clusters themselves,
during their formation period, are an important factor in determining
the observed size distribution.  As \citet{jor05} remark, ``The form of the
distribution ... should serve as a useful constraint for models of GC formation ...
any viable picture of star formation in clusters should produce an observed
size distribution that is consistent with the form of [the SDF]''.

A more well known feature of GC systems within
galaxies in general is their mass (luminosity) distribution (GCLF), which
is also a near-universal function relatively insensitive to host
galaxy size or environment.  The GCLF can be modelled as the outcome
of a power-law-like initial mass function coupled with many Gyr of
dynamical evolution, which preferentially removes the low-mass and low-density
clusters and reduces the IMF to the peaked Schechter-like form seen today
\citep[for only a handful of the dozens of papers discussing the 
mass distribution function and its secular evolution, see][]{baum03,whi07,jor07,mcl08b,gie08,kru09}.
By contrast, the origin of the linear size distribution for GCs
has received less attention.  But the state of development of
both observations and models is now reaching the point where some
paths to understanding it are opening.

As stated above, the GC effective radius remains nearly invariant over 
the normal dynamical evolution of the cluster\footnote{A \emph{proviso} to
this statement is that $r_h$ will grow slowly after core collapse, which
typically occurs after about 20 relaxation times; see, e.g., Trenti et al.~2007
among many modelling papers.} -- much more so than the
total mass of the clusters, which decreases by factors of 3 or more over a Hubble time
as it loses stars through the slow processes of tidal stripping and evaporation.
Thus $r_h$ gives us an unusually direct glimpse of its characteristic size
much closer to its origin.  However, extrapolating the current effective radius
all the way back to its protocluster epoch is highly unlikely to be correct.
A crucial stage in its evolution occurs just after formation, when
it consists of a mixture
of gas and stars in proportions determined by the star formation efficiency (SFE).
Especially over its first $\sim 30$ Myr, a cluster experiences rapid 
mass loss due to the energy output from its massive stars, including
UV radiation, supernova ejection, stellar winds, 
and even external dynamical heating.  These effects expel the gas and drive an internal
expansion of the cluster.  For the extreme case of low SFE and instantaneous gas loss
this stage would lead to rapid dissolution of the cluster into the field.  
For bound star clusters, however, both observations \citep{mac03,bas08} and theory 
\citep[see, e.g.][for a thorough overview]{baum07} show that for plausible SFEs,
the cluster expands typically by a factor of $2 - 5$ over this crucial
initial stage.  An empirical expression
for the growth of core radius with age for real clusters \citep{bas08} is 
$r_{core} = 0.6 {\rm ln} \lbrace (\tau/10^6y) - 0.25 \rbrace$.
After $\sim 10^8$ years, the cluster settles into
its long-term phase of slower internal dynamical evolution and 
stellar evaporation.

In addition, the observational evidence so far
\citep{mac03,bas08}, though admittedly still sketchy,
indicates that the extremely young clusters show a smaller
spread of core radii than the older ones.
It is therefore tempting to see the shape of the
present-day SDF (see again Figure \ref{rh_histo})
as the result of \emph{cluster-to-cluster
differences in the star formation efficiency}, starting from a population
of protocluster cores that began with rather similar sizes.  
The single-peaked but slightly asymmetric shape of the size distribution
is strongly reminiscent of a normal probability distribution that has
passed through a nonlinear transformation. In this case, the
transformation is the \emph{conversion of a given SFE to a radial expansion factor}.

In this view, the peak frequency at $r_h \simeq 2-3$ pc would simply 
represent protoclusters which experienced the most common (i.e. most probable) SFE.  
Smaller values of the initial SFE would lead to clusters with a larger present-day size, and
the very largest clusters would be ones with SFEs not much above the minimum level required
to remain bound.  At the opposite extreme, protoclusters with unusually high SFE
would suffer only very small expansion and end up at the minimum observed size
of $\sim 1$ pc.  Said differently, the ``rise point'' of the SDF at
$r_h \simeq 1$ pc should therefore be close to the typical initial size of the protoclusters.

The actual mean SFE within
any one protocluster should in principle be determined by several factors
(e.g. local temperature, pressure, turbulence, degree of initial
mass segregation), and thus 
could vary stochastically from one core to another.
However, the mean SFE is expected \citep[cf.][]{lada03} to be near 30\%
for dense star-forming regions that will give rise to massive clusters.
This mean SFE already permits an independent estimate of the amount of radial
expansion to be expected, from analytical energy arguments:  
in the limit of slow, adiabatic expansion during gas expulsion, the
product (cluster mass $\times$ radius) stays constant 
\citep{hills80}, which yields an expansion factor of $(SFE)^{-1} \simeq 3.3$.  This
suggests in turn that the initial sizes of the protoclusters should
be near $\sim 0.8$ pc to produce a present-day median around 2.5 pc,
consistent with the observations cited above.

The models by \citet{baum07} (BK07) allow us to explore some simple
simulations of the SDF a bit further
along the lines outlined above.  BK07 use three critical
factors determining the final expansion ratio $ER = r_h(final)/r_h(initial)$:
(a) the SFE in the protocluster, (b) the initial cluster size
relative to its tidal radius $(r_h/r_t)$, and (c) the ratio of the ``mass loss time'' $\tau_M$
over which gas is expelled relative to the internal crossing time $t_C$.
As argued above, we expect the protocluster core to be
typically $\sim 0.5 - 1$ pc in size.  This
level is far smaller than the typical tidal radius of $\ga 50$ pc for a
massive globular cluster,
so for our purposes we assume a small $(r_h/r_t) < 0.03$ (see Figure 4 of
BK07).  In addition, the very large amount of dense gas present in
the core during star formation will be capable of absorbing and thermalizing
the SNe ejecta and stellar winds, preventing instantaneous mass loss
\citep[see also][]{bailin09}.  In these massive protoclusters as well,
the crossing time $t_C \sim 10^5$ y is less
than the gas expulsion time \citep{goo97,baum08}.
We therefore use the BK07 models for 
the two representative cases $\tau_M = t_C$ (moderately rapid but not instant
gas expulsion) and $\tau_M = 10 t_C$ (slow, near-adiabatic gas expulsion).

It should be noted that we do not expect these setup arguments to carry over identically for
\emph{low-mass} clusters.  For these, the smaller tidal radius
will mean more rapid escape of stars during the expansion phase
as well as more rapid expulsion of gas \citep{baum08},  both of which put
the early evolution into a different regime of the model parameter space.
At masses well below the normal GC range, a far higher fraction of
the protoclusters will not survive this initial mass-loss stage.

With these assumptions about the tidal limit and mass loss time, the 
dispersion in the resulting cluster sizes \emph{starting from a fixed
initial core size} will be due simply to a dispersion in the SFE's
from one cluster to another.  To model the SFE in a simple way,
we assume that it follows a Gaussian distribution with a mean of $E_0$ and 
a standard deviation of $\sigma_E$.  In general, this approach resembles
the grid of simulations by \citet{baum08} drawing from the same models. However, their work
was directed towards exploring the trends of mean cluster mass and radius
as functions of galactocentric distance.  Here, we concentrate on attempting
to reproduce the detailed shape of the SDF itself.

Using the BK07 model grid, 
we find that the relation between the formation
efficiency and the expansion ratio $ER$ can be well approximated by
$ER \simeq 0.55 (SFE)^{-1.6}$ for the rapid case $\tau_M = t_C$.
As already noted above \citep[cf.][]{hills80}, 
$ER = (SFE)^{-1}$ for the slow-expulsion case $\tau_M = 10 t_C$.
We then run Monte Carlo realizations with these assumptions,
and vary $E_0, \sigma_E$, and the initial cluster size $r_h(init)$
to find simulations that match the real distributions.
A final bit of input is to convert the
initial cluster radius $r_h$ (three-dimensional) to the projected $r_h$
that we observe with the correction factor $r_h(proj) = 0.73 r_h(3D)$.

\begin{figure}
\includegraphics[angle=0,width=0.5\textwidth]{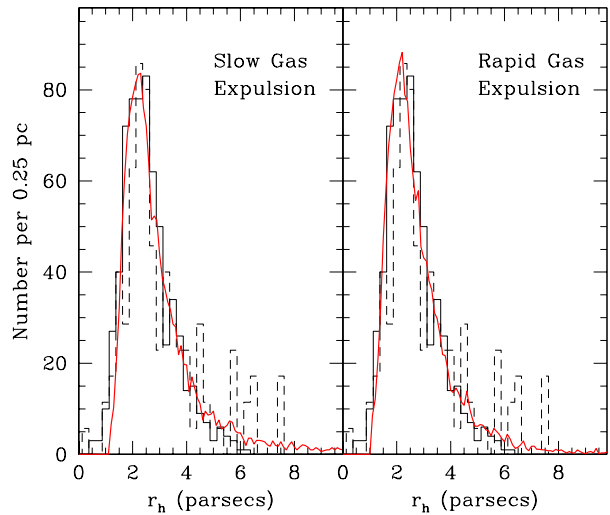}
\caption{The scale size distributions for the globular clusters
in M104 (solid histogram) and the Milky Way (dashed histogram), 
repeated from Figure \ref{rh_histo}.  In the \emph{left panel},
the solid red line shows a simulated $r_h$ distribution for
model clusters with slow gas expulsion and a range of star formation
efficiencies averaging $SFE = 0.28, \sigma_E=\pm0.11$
as described in the text.  In the \emph{right panel}
the red line shows a simulated distribution for rapid gas loss
and a mean $SFE = 0.30, \sigma_E=0.07$.
}
\label{sims}
\end{figure}

Figure \ref{sims} shows the results of two sample runs, matched
to the data shown previously for M104 and the Milky Way.
Each simulation is the result of generating $10^4$ clusters at random
following the prescriptions described above.  The left panel shows
the ``slow expulsion'' case for the parametric values $E_0 = 0.28$,
$\sigma_E = 0.11$, and
$r_h(init) = 1.0$ pc.  The right panel shows the ``rapid expulsion''
case for $E_0 = 0.30$, $\sigma_E = 0.07$, and $r_h(init) = 0.9$ pc.
Both the simulations and the Milky Way distribution have been
normalized to the total population of the M104 system.
Both models match the real data encouragingly well.  The three
important features are the rapid ramp-up in numbers of clusters
starting at $r_h \simeq 1$ pc; the moderately broad peak
near 2.5 pc; and the long but low-amplitude tail extending
to much larger radii.  

A limitation of the present discussion is obviously that we have
not properly included the long-term effects of dynamical evolution on the
SDF.  These effects would progressively trim the SDF that emerged
shortly after the early gas-loss phase, gradually removing low-mass
or very low-density objects.  For this reason, our SDF models which predict a few more clusters
at larger radii than in the present-day data are probably not a cause
for serious worry, because these large-radius, low-density clusters
are among the ones that would preferentially get destroyed by
dynamical evolution over the subsequent Hubble time.
The high$-r_h$ tail predicted by the models
should therefore be only an upper limit to the present-day distribution.

\citet{dac09} have collected the recent observations for
the SDF particularly in dwarf galaxies, and develop intriguing
evidence that the SDF in total may be {\sl bimodal}.  The
``normal'' mode peaked at $r_h \simeq 2.5-3.0$ pc is the
more well populated of the two, but there is a second mode peaked
near $r_h \simeq 8$ pc.  There are no obvious observational selection
effects that would bias discoveries (or size measurements) against
cluster in the intermediate-size range around 5 pc between the
two modes, so the reality
of the bimodality must be taken seriously.  Da Costa et al. propose
that these may belong to a second mode of cluster formation more
prevalent in the weaker potential wells of the dwarfs.

\begin{figure}
\includegraphics[angle=0,width=0.5\textwidth]{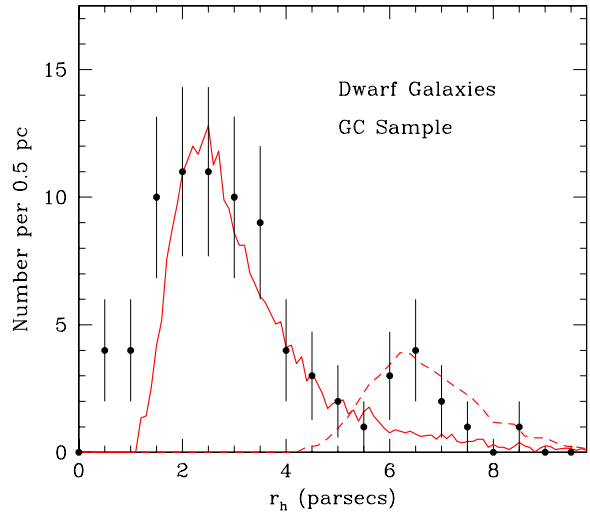}
\caption{The scale size distributions for the sample of globular clusters
in dwarf galaxies, from \citet{georgiev08} as described in the text.
The solid (red) line shows a simulated model distribution 
for the rapid gas expulsion
case with $E_0 = 0.32$, $\sigma_E = 0.08$, $r_h(init) = 1.15$ pc.
The dashed line shows a comparable simulation but now with
$E_0 = 0.30$, $\sigma_E = 0.03$, and $r_h(init) = 2.4$ pc.
}
\label{simdwarfs}
\end{figure}

Figure \ref{simdwarfs} shows a match of our model to the old-GC
sample compiled from several dwarf galaxies by \cite{georgiev08}
described above, where we attempt to match the bimodal distribution
noted by Da Costa et al.  The normal mode (solid line) is the fast-expulsion model
for $E_0=0.32$, $\sigma_E=0.08$, and $r_h(init)=1.15$ pc,
while the ``extended'' upper mode (dashed line) assumes
$E_0=0.30$, $\sigma_E=0.03$, and $r_h(init)=2.4$ pc.
Reproducing the upper mode requires a distinctly larger
initial protocluster radius, though interestingly, it needs to
be larger by only a factor of two.  The formal value of $\sigma_E$ is
quite a bit lower than for the normal mode, but once again
it is less clear how significant this is, because clusters that
ended up with \emph{very} large radii $r_h \ga 15$ pc might 
not have survived except (as also discussed by Da Costa et al.)
under the most favorable circumstances.
Minor variations around these
combinations of parameters can be found that give
similar fits, and the present discussion should be taken
only as illustrative.

\begin{figure}
\includegraphics[angle=0,width=0.5\textwidth]{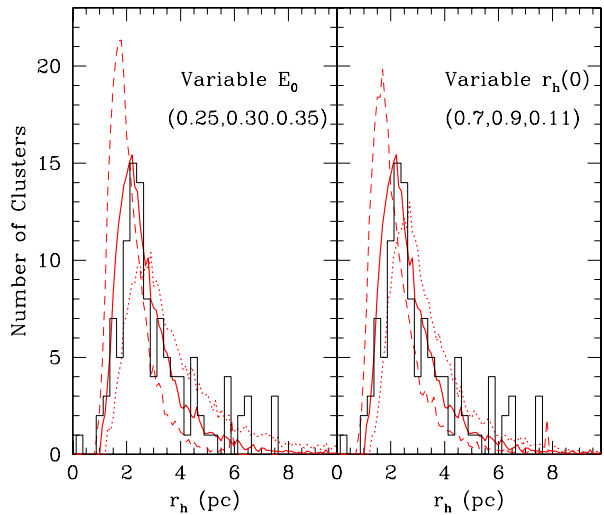}
\caption{\emph{Left panel:} The solid histogram shows the size distributions for the 
Milky Way globular clusters.  The red solid line is the ``best-fit'' model
for rapid gas expulsion ($E_0 = 0.30, \sigma_E = 0.07, r_h(init)= 0.9$ pc)
while the dashed and dotted lines show models for $E_0 = 0.35$ and $0.25$.
\emph{Right panel:} Same as in previous panel, but here the 
red dashed and dotted lines show models for $r_h(0) = r_h(init)=0.7$ and 1.1 pc.
}
\label{simrh}
\end{figure}

Lastly, Figure \ref{simrh} shows the sensitivity of the model
SDF to changes in both the basic SFE and the initial core size $r_h(init)$.  The same model
fit as in Figure \ref{sims}b for relatively rapid expulsion (solid line)
is shown ($E_0=0.30, \sigma_E=0.07, r_h(init)=0.9$)
along with the Milky Way cluster data.  The dashed and dotted
lines in the left panel show the same model but now with only 
the mean SFE changed ($E_0 = 0.35, 0.25$).  
The right panel shows models for changes in only the initial radius 
($r_h(init) = 0.7, 1.1$).  These four outlying models clearly fail to match the data. 
The implications are that the average starting conditions are
constrained to rather narrow ranges, $\pm0.05$ or less in $E_0$ and  $\pm0.1$ pc 
or less in $r_h(init)$. To a large extent, variations in $r_h(init)$ can be
traded off with variations in $E_0$ to produce the final range of observed sizes.

In rougher terms, the appropriate ranges for the initial conditions can be understood
physically as follows.  If $E_0$ falls below $0.20-0.25$,
then very few clusters survive the gas expulsion phase at all.
At the opposite end, values of $E_0$ higher
than about 0.4 mean that most clusters experience too little expansion
to reproduce the peak at 2.5 pc or to fill up the high$-r_h$ tail.
For the initial size, values of $r_h(init)$ bigger than $\sim 1$ pc or
less than $\sim 0.7$ pc drive the present-day peak well above or
below the characteristic 2.5-pc level that we need, and also fail to
match the observed absolute range in the SDF.  Finally, the third fitting parameter
$\sigma_E$ is essentially used to fine-tune the SDF peak and dispersion,
once $E_0$ and $r_h(init)$ have been put into the right range.  

These sample realizations are undoubtedly oversimplified, and do 
not by any means represent a thorough exploration
of the parameter space of the models.  This approach also
ignores other potentially interesting effects such as external interactions
with other clouds; mergers of young clusters; clumpiness and substructure
within a protocluster; or primordial mass segregation.  All of these could affect 
the early structural evolution \citep[e.g.][]{sce07,baum08b,mar08}.

Our main point, however, is that it seems possible to understand the key features of
the globular cluster size distribution 
with a relatively simple set of assumptions, and with quite plausible
fiducial values for the range of star formation efficiencies and
initial protocluster sizes.  Various arguments in the literature 
already favor a mean SFE of around 30 percent for star clusters
\citep[e.g.][]{lada03} and an initial size less than 1 pc \citep{bas08}.
However, there are far fewer avenues to quantitative estimates for the
\emph{range} of SFE's and initial scale sizes that typified 
the formation regions of massive star clusters.  The detailed shape of the present-day
size distribution appears to be one such method.

\section{Bimodality and the Mass/Metallicity Relation}

M104 was one of the first galaxies in which the intriguing correlation
between mean luminosity and color along the blue sequence
was found (Paper I).  This trend acts in the sense that the most luminous clusters
become slightly but progressively more metal-rich, and thus corresponds
to a \emph{mass/metallicity relation} (MMR).  The original discovery
papers used samples of data from large elliptical galaxies 
\citep{har06,str06,mie06}, but the M104 data indicated
that it might extend to disk systems as well.  The first round of
papers gave different results for the detailed shape of the MMR, but
more recent data analysis and discussions \citep{harris09,harris09b,peng09,cock09} indicate
more of a consensus emerging around the view that the blue sequence is nearly
vertical for lower luminosities ($L \la 5 \times 10^5 L_{\odot}$)
but gradually slants more toward the red going up to higher luminosities.
Equally intriguingly, no such MMR seems to affect the red sequence,
which keeps the same mean metallicity at all luminosities.

The most effective physical interpretation so far is based on
some form of self-enrichment during a cluster's formation stage
\citep{str08,bailin09}.  A protocluster of mass $\ga 10^7 M_{\odot}$ or more,
within a core of $\sim 1$ pc, will be able to hold back enough of its first round
of SN-enriched material to enrich the still-forming low-mass stars, thus
giving the entire cluster a higher mean metallicity than the pre-enriched
level it started with.  This extra self-enrichment will be much less
noticeable along the red sequence because its protocluster gas is an order
of magnitude higher in heavy-element abundance than the blue sequence.

We have used our new photometry of the M104 clusters to re-investigate
the existence of an MMR within this massive disk galaxy.  As noted
above, all the $BVR$ total magnitudes are individually corrected for
the scale sizes of the clusters and thus are free of any aperture-size
or PSF-fitting effects that might depend on luminosity.  
The first test is to measure any correlation of mean color with
luminosity along both sequences.  To do this, we use the $(B-R)$ data
as the most metallicity-sensitive of the three possible color indices
we could define, and then divide the data into half-magnitude bins
by $R$ magnitude.  The color distribution is then put into the
fitting routine RMIX \citep{weh08,harris09} and the best-fit bimodal
Gaussian distributions are found in each independent interval.  
In this way, we do not assume any particular form for the MMR
along either sequence.

\begin{table*}
\caption{Bimodal Fits to the Color Distributions}
\label{rmixtab}
\begin{tabular}{llrccc}
\hline
$R$ Range & $\langle R \rangle$ & n & $\mu_1$ & $\mu_2$ & Blue Fraction \\
\hline
22.5-23.5 & 22.92&  95& $1.118 \pm 0.013$ & $1.479 \pm 0.015$&$0.434 \pm 0.054$ \\
22.0-22.5 & 22.21& 103& $1.121 \pm 0.029$ & $1.448 \pm 0.031$&$0.463 \pm 0.102$ \\
21.5-22.0 & 21.74& 131& $1.162 \pm 0.016$&  $1.452 \pm 0.022$&$0.489 \pm 0.070$ \\
21.0-21.5 & 21.29& 116& $1.169 \pm 0.022$&  $1.449 \pm 0.035$&$0.507 \pm 0.104$ \\
20.5-21.0 & 20.77&  82& $1.175 \pm 0.013$ & $1.494 \pm 0.017$&$0.549 \pm 0.059$ \\
20.0-20.5 & 20.28&  55& $1.214 \pm 0.017$ & $1.470 \pm 0.027$&$0.548 \pm 0.087$ \\
19.5-20.0 & 19.78&  36& $1.178 \pm 0.018$ & $1.499 \pm 0.028$&$0.561 \pm 0.088$ \\
19.0-19.5 & 19.32&  17& $1.258 \pm 0.032$ & $1.518 \pm 0.047$&$0.522 \pm 0.153$ \\
18.0-19.0 & 18.68&  16& $1.278 \pm 0.022$ & $1.520 \pm 0.061$&$0.733 \pm 0.124$ \\
\hline
\end{tabular}
\end{table*}

\begin{figure}
\includegraphics[angle=0,width=0.5\textwidth]{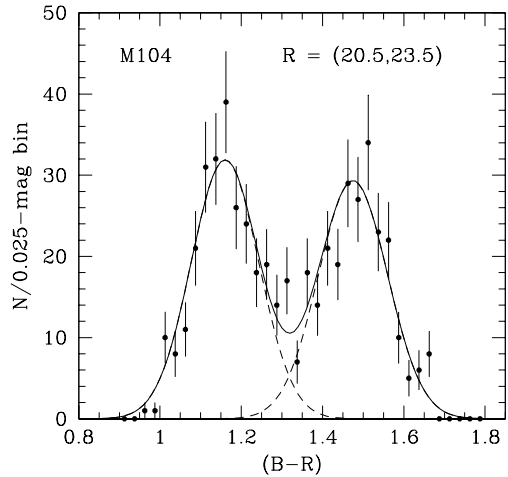}
\caption{Histogram of $(B-R)$ colors for the M104 globular clusters
in the magnitude interval $20.5 < R \leq 23.5$ (solid points
with errorbars).  The solid curve shows the best-fit bimodal Gaussian,
with the two individual modes shown as dashed lines.
}
\label{cfit}
\end{figure}

\begin{figure}
\includegraphics[angle=0,width=0.5\textwidth]{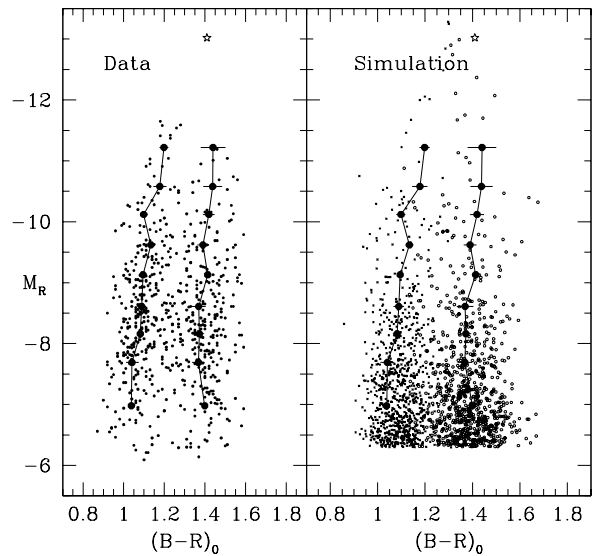}
\caption{\emph{Left panel:}  Color-magnitude data for the globular clusters
in M104, from Figure \ref{cmdraw} and transformed to absolute magnitude $M_R$
versus intrinsic color $(B-R)_0$.  Along each sequence the mean points in
half-magnitude intervals (Table \ref{rmixtab}) are plotted as the connected
large dots.  The large star at top marks the UCD.
\emph{Right panel:}  A simulated set of data from a bimodal metallicity
distribution, as described in the text.  At high mass the clusters are
noticeably affected by self-enrichment.
}
\label{cmd2}
\end{figure}

The mean points derived from these objective fits are listed in Table \ref{rmixtab}
and plotted in Figure \ref{cmd2}.  In principle, RMIX can solve for five
free parameters:  the mean colors $\mu_1, \mu_2$ of the blue and red modes;
their Gaussian dispersions $\sigma_1, \sigma_2$; and the proportion $p_1$
(or $p_2 = 1-p_1$) that the blue (red) mode makes up of the total population.
In practice, if the number of datapoints in the bin is less than about 50,
the solutions need to be partially constrained for convergence; here, we choose
in such cases to fix the dispersions $\sigma_{1,2}$ because these do not
change noticeably along the sequence and we are primarily interested in
tracing the mean colors themselves.  By using the total data over
$20.5 < R < 22.0$ we find $\langle \sigma_1 \rangle = 0.070 \pm 0.006$
and $\langle \sigma_2 \rangle = 0.096 \pm 0.010$.  
The color distribution in $(B-R)$ for this galaxy defines
a remarkably clear bimodal histogram (Figure \ref{cfit}),
and the double-Gaussian model provides an accurate fit.

As can be seen from Table \ref{rmixtab}, the blue and red clusters make
up nearly equal proportions of the total population; the $p_1(blue)$ ratio
increases steadily toward higher luminosity and the blue sequence reaches
a bit higher at the top end.  As expected from both the previous literature
and the self-enrichment model, the red sequence does not
show a significant change in mean color with luminosity (that is, it has a
``zero'' MMR).  By contrast, the blue sequence shows a clear slope
toward the red that is nearly linear in form.  An unweighted linear
fit to the mean points in Table \ref{rmixtab} gives a highly
significant slope $\Delta (B-R) / \Delta M_R = -0.037 \pm 0.005$,
almost identical with what was derived in Paper I.
In terms of heavy-element abundance $Z$ this slope corresponds
to a scaling $Z \sim L^{0.29 \pm 0.04}$ (see below).

The slope of the linear fit
becomes progressively less significant as the higher-luminosity
bins are removed. For example, a direct fit of $(B-R)$ versus $R$ for the
clusters with $M_R > -8$ and $(B-R)_0 < 1.20$ gives
$\Delta(B-R) / \Delta M_R = -0.013 \pm 0.014$.  
Thus we cannot rule out the possibility that the blue sequence may
become more nearly vertical at lower luminosities.
Nevertheless, we
confirm the basic trend found in Paper I that the blue sequence does
show an MMR.  In addition, the metallicity scaling agrees extremely well with
the mean slopes $Z \sim L^{0.3}$ found for several giant E galaxies
in the recent studies by \citet{harris09}, \citet{harris09b}, and \citet{cock09},
and gives some additional support to the idea that the MMR may be
a near-``universal'' phenomenon which requires a broad-based
physical explanation independent of galaxy type.

As a more direct comparison between theory and data, we show in
Figure \ref{cmd2}b the same mean points from the color-magnitude diagram,
but now superimposed on a simulation drawn from the \citet{bailin09}  theory.
In this model, the clusters are assumed to form from dense protoclusters
of mass $M_0$, initial size $r_0$, and with star formation efficiency $E_0$.  
The protocluster gas has a pre-enriched metallicity level [m/H]$_0$,
which is then enriched further by some fraction of the 
first SNe that go off in the emerging cluster.  After its formation
stage, the early gas expulsion leaves the cluster with a mass $M_0 \times E_0$.
Then to take account of the longer-term dynamical mass loss that follows, we use the
conventional expression for a roughly constant mass loss rate applying
to two-body relaxation in tidally limited clusters, $M(t) \simeq M_0 - \mu_{ev} t$ 
\citep[e.g.][]{baum03,mcl08b}.  The actual mass loss rate $\mu_{ev}$ is expected
to be a function of cluster density; for example, \citet{mcl08b} derive
$\mu_{ev} = 1100 \rho_h^{1/2}$ $M_{\odot}$ Gyr$^{-1}$ 
where $\rho_h$ is in $M_{\odot}$ pc$^{-3}$.  The empirical evidence shows that
the characteristic density $\rho_h$ in turn increases
systematically with mass roughly as $\rho_h \sim M$, though it shows large
cluster-to-cluster scatter (see Figure 1a of McLaughlin and Fall, for example).
A fully detailed simulation of these effects is beyond the scope of our
discussion, but as a first-order representation of the mean trend, we use
$\mu_{ev} \simeq 5000 (M/10^5 M_{\odot})^{1/2}$ $M_{\odot}$ Gyr$^{-1}$,
which reasonably reproduces the mass loss rates in the 
N-body simulations by \citet{baum03} and more recent work of 
\citet{kru09} and 
Hurley (private communication).  To set the other parameters in the enrichment model,
for all clusters we use $E_0 = 0.3, r_0 = 1$ pc (Section 4 
above).  We also use a supernova conversion efficiency $f_{SN} = 0.3$  (that is, 
30\% of the SNe in the cluster formation sequence happen soon enough in the burst to 
help enrich the remaining gas).  

The initial (pre-enriched) values for the metallicities of each sequence are
chosen to match the two observed sequences, at $(B-R)_0 = 1.09$ (blue)
and 1.39 (red). (Note here that these are the dereddened colors, which we
use for the simulations and plot in Figure \ref{cmd2}.  The data in Table 2
are the directly observed colors with no reddening removed.)
Transformation between metallicity and color is determined by
$(B-R)_0 \, = \, 0.27 {\rm [Fe/H]} + 1.52$ \citep{bar00,can07}.  We assume
a mass-to-light ratio $(M/L)_R = 2$ and convert to absolute magnitude with
$M_R = 4.29 - 2.5 {\rm log} (L/L_{\odot})$.
Finally, the sequences are assumed to have intrinsic
dispersions in metallicity of $\pm 0.26$ dex (blue) and 0.36 dex (red).

With this input, we then construct 
Monte Carlo realizations of the entire GCS, a typical example of which is shown in Figure \ref{cmd2}b.
The initial mass distribution is drawn randomly from a simple power law form
$dN/dM_0 \sim M_0^{-2}$, and
the total number of simulated clusters is chosen to match the observed numbers
in M104 for $M_R \la -8$.   Because of our overly simplistic initial mass function
for the input clusters, the simulation overestimates the numbers both at very
high and low luminosity; however,
the important feature is the matchup in the mean position of each sequence.

We find that the self-enrichment model can reproduce the red and blue sequences
basically well.  On the blue sequence, the ``cut-on'' point where self-enrichment starts
clearly to dominate over the pre-enriched level is at $M_R \simeq -10$ ($M \simeq 10^6 M_{\odot}$), 
above which the sequence starts to slant more strongly to the red.  
At very high luminosity, both sequences converge towards an enrichment
level [m/H] $\simeq -0.4$ or $(B-R)_0 \simeq 1.4$, very 
near where the M104 UCD sits in the color-magnitude diagram.  

A second-order concern is the discrepancy between model 
and data for the low-luminosity range $M_R > -8$.
The real clusters in this range appear to be slightly more enriched (redder)
than their simulated counterparts.
The model, as it stands, firmly predicts a
``zero MMR'' at present-day masses below about $10^5 M_{\odot}$; even in their
more massive protocluster state, such clusters are simply
not massive or dense enough to hold back significant amounts of the enriched SNe ejecta.
In other words, the model provides a good quantitative explanation for the MMR at
the upper end of the sequences, but cannot explain any MMR that might continue to lower masses.
Thus \emph{if} the MMR genuinely extends to lower luminosities (which we take our
data to hint at, though not conclusively), either the model is incomplete or another
explanation must exist.

The main free parameters in the model are the SFE $E_0$, initial size $r_0$,
and SN conversion efficiency $f_{SN}$.
In principle, we could adjust these to force the blue cut-on point
to lower luminosities:  for example, making the SFE mass-dependent such that
$E_0$ is higher at \emph{lower} masses, or making $r_0$
smaller for the lower-mass clusters,
would raise the resulting enrichment at lower mass.  
We could also make $f_{SN}$ arbitrarily higher only for the lower-mass clusters,
but it is not clear what might produce such a counterintuitive effect.
Decreasing $r_0$ to 0.5 pc or increasing $f_{SN}$ to 1.0
have about the same effect of lowering the cut-on point by less than a factor
of two, and combining both effects would reduce it to $4 \times 10^5 M_{\odot}$.
However, changes of this type away from their nominal values of
$E_0 = 0.3$ and $r_0 \simeq 1$ pc would dramatically alter the resulting
SDF and are thus ruled out by the analysis presented in Section 4 above.
In addition, they would increase the self-enrichment at higher luminosity
even further, and destroy the matchup with the real data for $M > 10^6 M_{\odot}$.

For the present time, we see no reason to discard the basic approach of
the self-enrichment model, which is quite effective at matching the
central features of the MMR particularly for the high-luminosity clusters.
The present study should be viewed as only an indication that further
interesting tests may rely on the detailed features of the GC metallicity
distribution over the entire run of luminosities.

\section{Summary}

In this study we have used a mosaic of HST/ACS images of the Sombrero
galaxy (M104) to make new measurements of its large globular cluster
population.  We obtain new $BVR$ photometry and effective-radius measurements
for a total of 652 clusters.  Our findings include the following:

\noindent (1) The individual cluster profiles have been fitted with
King and Wilson dynamical models to derive their effective (half-light)
radii $r_h$.  A series of tests of the data with different
fitting models and different definitions of the stellar point-spread function
show that the internal precision of the half-light radii is typically $\pm0.4$ pc.  

\noindent (2) The distribution function of the effective radii (SDF) has a peak near $r_h = 2.4$
pc and an asymmetric tail to larger radii, which to first order is 
remarkably similar to the SDFs found for other galaxies, including the Milky
Way, giant E galaxies, and several kinds of dwarf galaxies.  We find second-order trends with cluster
metallicity and spatial location (the clusters are slighter bigger on
average at lower metallicity, or larger galactocentric distance).
Both of these trends also resemble what has been found in other
galaxies in several other recent studies.  However, closer comparisons
with other data in the literature show that the metallicity-based size
difference may not be as large for GCs in disk galaxies as in E galaxies.

\noindent (3) The M104 GCs define very much the same regions in
the structural Fundamental Plane as in other, more well resolved systems
including the standard Milky Way.  We use the half-light surface brightness
$I_{V,h}$ and binding energy $E_b$ to demonstrate that clusters become
preferentially more extended at larger galactocentric distances or
low luminosities.  As long as the cluster half-light radii can be
measured accurately, these quantities can be used in still more
distant systems such as the giant ellipticals in Virgo and elsewhere.

\noindent (4) We have explored a simple framework for the physical origin
of the globular cluster scale-size distribution.  
We assume that GCs start as dense,
massive protoclusters which form stars at a certain efficiency and then
expand to radii near their present-day sizes during a rapid initial stage of
mass loss and residual gas expulsion.  Using models by \citet{baum07},
we find that the observed
SDF can be successfully and closely matched if we assume that (a) the protoclusters
began with scale sizes $\simeq 0.8-0.9$ pc; (b) the mean star formation
efficiency was $\simeq 0.30$ but with stochastic cluster-to-cluster variations on
the order of $\pm 0.08-0.15$; and (c) the gas expulsion time was at least as
long as the internal crossing time.

\noindent (5) The color-magnitude distribution for the GCs shows a clearly
defined, classic bimodal form with nearly equal numbers of metal-poor and metal-rich
clusters.  Detailed bimodal fitting of the $(B-R)$ colors shows that 
the blue, metal-poor sequence exhibits a well determined mass/metallicity relation
(MMR), becoming slightly but steadily redder towards higher luminosity.
The red sequence is vertical over the entire luminosity range.
Detailed comparisons with the self-enrichment model of \citet{bailin09}
show that a close first-order match to the data can be obtained with
the same choice of star formation efficiency and protocluster size
that are required to successfully model the SDF.  However, a potential
problem for the model may lie in whether or not the MMR extends further
down the blue sequence below the point where self-enrichment can
be expected to work.  This will be an intriguing area for future work.

\section*{Acknowledgements}

We are especially grateful to
Dean McLaughlin for making his profile fitting code publicly available,
and for a critical and constructive reading of the text.
We also thank Nate Bastian and Mark Gieles for helpful guidance.
WEH is pleased to acknowledge support from the Natural Sciences and Engineering
Research Council of Canada and the Killam Foundation of the Canada Council.
DAF and LRS thank the Australian Research Council for their financial
support.

\bsp

\label{lastpage}

\end{document}